\newif\ifsubmode
\submodefalse

\newif\ifprintfig
\printfigtrue

\ifsubmode
  \documentstyle[12pt,aasms4,epsf]{article}
  \received{}
  \accepted{}
  \journalid{}{}
  \articleid{}{}
\else
  \documentstyle[11pt,aaspp4,epsf]{article}
  \slugcomment{{\it submitted to the Astrophysical Journal}}
\fi

\lefthead{Cretton \& van den Bosch}
\righthead{Evidence for a massive BH in the S0 galaxy NGC 4342}

\newcommand{\etal}{{et al.~}}

\newcommand{\gta}{\gtrsim}

\newcommand{\kms}{\>{\rm km}\,{\rm s}^{-1}}

\newcommand{\Msun}{\>{\rm M_{\odot}}}
\newcommand{\Lsun}{\>{\rm L_{\odot}}}

\begin{document}

\title{Evidence for a massive BH in the S0 galaxy NGC 4342}
       
\author{Nicolas Cretton}
\affil{Leiden Observatory, PO Box 9513, 2300 RA  Leiden, The
       Netherlands}

\author{Frank C. van den Bosch\altaffilmark{1}}
\affil{Department of Astronomy, University of Washington, Seattle, 
       WA 98195, USA\altaffilmark{2}}

%%%%%%%%%%%%%%
% Additional affiliations
%%%%%%%%%%%%%%

\altaffiltext{1}{Hubble Fellow}
\altaffiltext{2}{Previously at Leiden Observatory, PO Box 9513, 2300 
RA Leiden, The Netherlands.}

%%%%%%%%%%%%%%%
% Start the abstract on a fresh page
%%%%%%%%%%%%%%%

\ifsubmode\else
\clearpage\fi

%%%%%%%%%%%%%%%
% Use a small baselineskip, unless in submission mode.
%%%%%%%%%%%%%%%

\ifsubmode\else
\baselineskip=14pt
\fi

%%%%%%%%%%%%%%%
% Abstract
%%%%%%%%%%%%%%%

\begin{abstract}
  We present axisymmetric dynamical models of the edge-on S0 galaxy
  NGC~4342.  This small low-luminosity galaxy harbors, in addition to
  its outer disk, a bright nuclear stellar disk.  A combination of
  observations from the ground and with the Hubble Space Telescope
  (HST) has shown that NGC~4342 rotates rapidly and has a strong
  central increase in velocity dispersion.
 
  We construct simple two-integral Jeans models as well as fully
  general, three-integral models.  The latter are built using a
  modified version of Schwarzschild's orbit-superposition technique
  developed by Rix \etal and Cretton \etal These models allow us to
  reproduce the full line-of-sight velocity distributions, or
  `velocity profiles' (VPs), which we parameterize by a Gauss-Hermite
  series. The modeling takes seeing convolution and pixel binning into
  account.

  The two-integral Jeans models suggest a black hole (BH) mass between
  $3$ and $6\times 10^8 \Msun$, depending on the data set used to
  constrain the model, but they fail to fit the details of the
  observed kinematics. The three-integral models can fit all
  ground-based and HST data simultaneously, but only when a central BH
  is included.  Models without BH are ruled out to a confidence level
  better than $99.73$ per cent. We determine a BH mass of
  $3.0^{+1.7}_{-1.0} \times 10^8 \Msun$, where the errors are the
  formal 68.3 per cent confidence levels. This corresponds to 2.6 per
  cent of the total mass of the bulge, making NGC~4342 one of the
  galaxies with the highest BH mass to bulge mass ratio currently
  known.

  The models that best fit the data do not have a two-integral
  phase-space distribution function. They have rather complex
  dynamical structures: the velocity anisotropies are strong functions
  of radius reflecting the multi-component structure of this galaxy.

  When no central BH is included the best fit model tries to fit the
  high central velocity dispersion by placing stars on radial orbits.
  The high rotation velocities measured, however, restrict the amount
  of radial anisotropy such that the central velocity dispersion
  measured with the HST can only be fit when a massive BH is included
  in the models.
\end{abstract}

%%%%%%%%%%%%%%%
% Keywords
%%%%%%%%%%%%%%%


\clearpage

%%%%%%%%%%%%%%%
% Beginning of main text
%%%%%%%%%%%%%%%

\section{Introduction}

Several lines of evidence suggest that active galactic nuclei (AGNs)
are powered by accretion onto a super-massive black hole (BH)
(Lynden-Bell 1969; Rees 1984). The much higher volume-number density
of AGNs observed at redshift $z \approx 2$ than at $z=0$, suggests
that many quiescent (or `normal') galaxies today must have gone
through an active phase in the past, and therefore harbor a massive
BH as well. Such a BH will significantly influence the dynamics of the
galaxy inside a radius of influence, $r_{\rm BH} = G M_{\rm
BH}/\sigma^2$, where $\sigma$ is a characteristic velocity dispersion
of the stars in the center. In particular, hydrostatic equilibrium
requires that the rms velocities of the stars surrounding a massive BH
follow an $r^{-1/2}$ power-law (Bahcall \& Wolf 1976; Young 1980).

Since the late 70s, combined imaging and spectroscopy of the central
regions of galaxies has suggested that massive BHs should be present
in a number of early-type galaxies (see Kormendy \& Richstone 1995 for
a review). Conclusive dynamical evidence for the presence of a central
BH requires that a model with a BH can fit all observations
(photometric and kinematic), and that no model without a BH can
provide an equally good fit. Such conclusive evidence can only be
inferred from observations that probe well inside the radius where the
BH dominates the dynamics. Up to a few years ago, most claimed BH
detections were based on observations with spatial resolutions of
similar size as the radii of influence of the inferred BH masses (Rix
1993). This, together with the limited amount of freedom in the models
used to interpret the data, has hampered an unambiguous proof for the
presence of these BHs (i.e., the observed kinematics could not be
confronted with all possible dynamical configurations without a BH).
Often spherical models were used even when the observed flattening was
significant. If the models were axisymmetric, the distribution
function (hereafter DF) was often assumed to depend only on the two
classical integrals of motion, energy and vertical angular momentum;
$f = f(E,L_z)$. This implies that the velocity dispersions in the
radial and vertical directions are equal (i.e., $\sigma_R =
\sigma_z$). It is well-known that strong radial anisotropy in the
center of a galaxy results in a high central velocity dispersion,
mimicking the presence of a massive BH (cf. Binney \& Mamon 1982).
Conclusive evidence for a BH therefore requires that one can rule out
radial anisotropy as the cause of the high velocity dispersions
measured, and models must thus be sufficiently general.

Recently two major breakthroughs have initiated a new era in the
search for massive BHs in normal galaxies. First of all, we can now
obtain kinematics at much higher spatial resolution (down to FWHM
$\sim 0.1''$), using specially-designed spectrographs, such as the
Subarcsecond Imaging Spectrograph (SIS) on the Canada-France Hawaii
Telescope, or the Faint Object Spectrograph (FOS) and STIS aboard the
HST. This allows us to probe the gravitational potential much closer
to the center, where the BH dominates the dynamics. Not only has this
improved the evidence for massive BHs in several old BH-candidate
galaxies (M31, Ford \etal 1998; M32, van der Marel \etal 1997, 1998;
M87, Harms \etal 1994, Macchetto \etal 1997; NGC~3115, Kormendy \etal
1996a; NGC~4594, Kormendy \etal 1996b), but it has also provided new
cases (M84, Bower \etal 1998; NGC~3377, Kormendy \etal 1998; NGC~3379,
Gebhardt \etal 1998; NGC~4261, Ferrarese, Ford \& Jaffe 1996;
NGC~4486B, Kormendy \etal 1997; NGC~6251, Ferrarese, Ford \& Jaffe
1998; and NGC~7052, van der Marel \& van den Bosch 1998).  Secondly,
the revolutionary increase in computer power has made it possible to
investigate a large number of fully general, three-integral models
based on the orbit-superposition method (Schwarzschild 1979). In the
past decade, this method has been used to build a variety of
spherical, axisymmetric and triaxial models (e.g., Schwarzschild 1982;
Pfenniger 1984; Richstone \& Tremaine 1984, 1988; Zhao 1996).  Levison
\& Richstone (1985), Richstone \& Tremaine (1985), and Pfenniger
(1984) showed how to include rotation velocities and velocity
dispersions as kinematic constraints.  More recently, Rix \etal (1997)
and Cretton \etal (1998) extended this modeling technique even further
by fitting to the {\it entire} velocity profiles (see also Richstone
1997).  Van der Marel \etal (1997, 1998) used this to build fully
general, axisymmetric models of M32, and showed convincingly that M32
harbors a massive BH.  Recent review papers on this rapidly evolving
field include Ford \etal (1998), Ho (1998), Richstone (1998), and van
der Marel (1998).

In many galaxies where the presence of a BH has been suggested, a
nuclear disk, seen close to edge-on, is present. These disks are
either in gaseous form (M84, M87, NGC~4261, NGC~4594, NGC~6251,
NGC~7052), or made up of stars (NGC~3115). It is easier to detect BHs
in edge-on systems with disks, where one can use both the measured
rotation velocities and the velocity dispersions to determine the
central mass density. It is therefore not surprising that BHs have
predominantly been found in galaxies with nuclear disks. Furthermore,
nuclear disks allow a good determination of the central mass density
of their host galaxies.  Gaseous disks have the advantage that their
kinematics can be easily measured from emission lines. Since gas in a
steady-state disk can only move on non-intersecting orbits, the
measured rotation velocities of a settled gas disk, in the equatorial
plane of an axisymmetric potential, correspond to the circular
velocities, $V_c(R) = \sqrt{R {\rm d}\Phi/{\rm d}R}$.  The rotation
curve of a nuclear gas disk therefore provides a direct measure of the
central potential gradient, and thus of the central mass density.
However, often the gas disks are not in a steady state; many show a
distorted morphology (e.g. M87, see Ford \etal 1994), and
non-gravitational motion, such as outflow, inflow or turbulence can be
present and complicate the dynamical analysis (e.g., NGC~4261, Jaffe
\etal 1996; NGC~7052, van den Bosch \& van der Marel 1995).
Nuclear {\it stellar} disks do not suffer from this, but have the
disadvantage that their kinematics are much harder to measure.  First
of all, the kinematics have to be determined from absorption lines
rather than emission lines, and secondly, the line-of-sight velocity
distributions, or velocity profiles (VPs), measured are `contaminated'
by light from the bulge component.  However, van den Bosch \& de Zeeuw
(1996) showed that with sufficient spatial and spectral resolution one
can resolve the VPs in a broad bulge-component and a narrow
disk-component. From these VPs the rotation curve of the nuclear disk
can be derived, providing an accurate measure for the central mass
density. Therefore, galaxies with an embedded nuclear disk (either
gaseous or stellar) observed close to edge-on are ideal systems to
investigate the presence of massive BHs.

In this paper we discuss the case of NGC~4342; a small, low-luminosity
($M_B = -17.47$) S0 galaxy in the Virgo cluster. The galaxy is listed
as IC~3256 in both the Second and Third Reference Catalogues of Bright
Galaxies, since in the past it has occasionally been confused with
NGC~4341 and NGC~4343 (see Zwicky \& Herzog 1966).  At a projected
distance of $\sim 30''$ SE of NGC~4342, a small galaxy is visible. It
is uncertain whether this is a real companion of NGC~4342 or whether
it is merely close in projection. HST images of NGC~4342 revealed both
an outer disk, as well as a very bright nuclear stellar disk inside
$\sim 1''$ (van den Bosch \etal 1994; Scorza \& van den Bosch 1998).
It is a normal galaxy, with no detected ISM (Roberts \etal 1991),
and with small color-gradients (van den Bosch, Jaffe \& van der Marel
1998, hereafter BJM98).  For its size and luminosity, it does however
reveal a remarkably large central velocity dispersion and a very steep
rotation-curve (see BJM98). Unfortunately, the spectral resolution of
the available kinematic data is insufficient to actually resolve the
VPs in disk and bulge components. In order to determine the central
mass density in NGC~4342, we thus have to construct dynamical models
of the entire system: bulge and disk components.  Here we present
simple two-integral Jeans models as well as fully general
three-integral models, and we provide evidence for the presence of a
central massive dark object (MDO) of $\sim 3\times 10^8 \Msun$.
Throughout this paper we assume the MDO to be a BH, but we discuss
alternatives in Section~7.2

In Section~2 we briefly discuss the data used to constrain the models
and in Section~3 we describe our mass model. In Section~4 we show the
results of some simple two-integral modeling, and we discuss its
shortcomings. Section~5 describes the general outline of the
three-integral modeling technique. In Section~6 we discuss
shortcomings of the velocity-profile parameterization used when
applied to dynamically cold systems, and present a modified approach.
The results of the three--integral modeling are discussed in
Section~7. Finally, in Section~8, we sum up and present our
conclusions. Throughout this paper we adopt a distance of 15 Mpc for
NGC~4342, consistent with the distance of the Virgo cluster (Jacoby,
Ciardullo \& Ford 1990).

\section{The data}

All data used in this paper are presented and discussed in detail in
BJM98. Here we merely summarize. 

\subsection{Photometric data}

BJM98 used the Wide Field and Planetary Camera 2 (WFPC2) aboard the HST 
to obtain $U$, $V$ and $I$ band photometry of NGC~4342. The
spatial resolution of these images is limited by the HST Point Spread
Function (FWHM $\sim 0.1''$) and the size of the pixels ($0.0455''
\times 0.0455''$).  The full field-of-view covers about $35'' \times
35''$. Figure~\ref{fig:mge} shows contour plots at two different
scales of the $I$-band image. The presence of the nuclear disk is
evident from the highly flattened, disky isophotes inside $1.0''$.

\subsection{Kinematic data}

Using the ISIS spectrograph mounted at the 4.2m William Herschel
Telescope (WHT) at La Palma, BJM98 obtained long-slit spectra of
NGC~4342 along both the major and the minor axis. The spectra have a
resolution of $\sigma_{\rm instr} = 9\kms$, and were obtained with a
slit width of $1.0''$ under good seeing conditions with a PSF FWHM of
$0.80''$ (major axis) and $0.95''$ (minor axis). After standard
reduction, the parameters $(\gamma,V,\sigma,h_3,h_4)$ that best fit
the VPs were determined using the method described in van der Marel
(1994). These parameters quantify the Gauss-Hermite (GH) expansion of
the velocity profile ${\cal L}(v)$ as introduced by van der Marel \&
Franx (1993):
\begin{equation}
\label{ghseries}
{\cal L}(v) = {\gamma \over \sigma} \alpha(w) \left( 1 + \sum_{j=3}^{4}
h_j H_j(w) \right) ,
\end{equation}
where
\begin{equation}
\label{weq}
w \equiv (v-V)/\sigma,
\end{equation}
and
\begin{equation}
\label{alphaweq}
\alpha(w) = {1 \over \sqrt{2 \pi}} {\rm e}^{-{1\over 2} w^2}.
\end{equation}
Here $v$ is the line-of-sight velocity, $H_j$ are the Hermite
polynomials of degree $j$, and $h_j$ are the Gauss-Hermite
coefficients.  The first term in equation~(\ref{ghseries}) represents
a Gaussian with line strength $\gamma$, mean radial velocity $V$, and
velocity dispersion $\sigma$.  The even GH-coefficients quantify
symmetric deviations of the VP from the best-fitting Gaussian, and the
odd coefficients quantify the asymmetric deviations.

%We have folded the kinematic WHT data by calculating the weighted mean
%of the data points on different sides of the nucleus.
We have averaged the kinematic WHT data at positive and negative
radii.  In this way we obtain sets of $(V,\sigma,h_3,h_4)$ at 19
different positions along the major axis and 8 along the minor axis.

BJM98 also obtained FOS spectra at 7 different aperture-positions, all
inside the central $0.5''$ of NGC~4342, using the circular
$0.26''$-diameter aperture (the {\tt FOS 0.3}-aperture). Due to the
limited signal-to-noise ratio ($S/N$) of these spectra, only
$(\gamma,V,\sigma)$ of the best-fitting Gaussian could be determined.
Rotation velocities and velocity dispersions along the major axis of
NGC~4342 for both the WHT (crosses) and FOS (solid dots) are shown in
Figure~\ref{fig:kinemdat}.  The central rotation gradient, as measured
with the FOS, is extremely steep ($V \sim 200 \kms$ at $0.25''$ from
the center). In addition, the velocity dispersion increases from $\sim
90 \kms$ at the outside (the `cold' outer disk) to $317 \kms$ in the
center as measured with the WHT.  The central velocity dispersion
increases to $418 \kms$, when observed at four times higher spatial
resolution with the FOS.

\section{The mass model}

We have used the Multi-Gaussian Expansion (MGE) method developed by
Emsellem, Monnet \& Bacon (1994, hereafter EMB94) to build a mass
model for NGC~4342. The method assumes that both the PSF and the
intrinsic surface brightness are described by a sum of Gaussians, each
of which has 6 free parameters: the center $(x_j,y_j)$, the position
angle, the flattening $q'_j$, the central intensity $I'_j$, and the
size of the Gaussian along the major axis, expressed by its standard
deviation $a'_j$.  The best-fitting parameters of the different
Gaussians are determined using an iterative approach in which
additional components are added until convergence is achieved (see
EMB94 for details on the method). This method is well suited for
complicated, multi-component galaxies such as NGC~4342.

We fitted the HST $I$-band PSF by a sum of 5 circular (i.e., $q'_j=1$)
Gaussians (see BJM98). Using this model PSF we derived the parameters
of the $N$ Gaussians describing the intrinsic surface brightness
(i.e., deconvolved for PSF effects) by fitting to the HST $I$-band
image of NGC~4342. We forced the $N$ Gaussians to have the same
position angle and center, which yields an axisymmetric mass model
(see below). Therefore, the model is described by $3N + 3$ free
parameters, which are simultaneously fit to the image. We achieved
convergence with $N=11$ Gaussian components. The results of the fit
are shown in Figure~\ref{fig:mge}, where we show contour plots of the
$I$-band image with superimposed contours of the convolved surface
brightness of the MGE model. The fit is excellent, except for a small
discrepancy at the outside. This is due to slight twisting of the
isophotes at large radii (see BJM98). Since our model is axisymmetric
, this cannot be modeled. Nevertheless, the discrepancy is
small, and is unlikely to affect our conclusions on the dynamics of
the central region.  The parameters of the different Gaussian
components are listed in Table~1.

The total luminosity of the MGE model in the $I$-band is $L_I = 3.57
\times 10^9 \Lsun$.  This yields $M_I = -19.86$. The absolute blue
magnitude of NGC~4342, at a distance of 15 Mpc, is $M_B = -17.47$
(Sandage \& Tammann 1981), and we thus find $B-I = 2.39$. This is
consistent with the colors of NGC~4342 presented by BJM98. They find
$U-V \approx 1.5$ and $V-I \approx 1.3$.  We thus derive $B-V \approx
1.09$, in good agreement with the average value for early-type
galaxies (Faber \etal 1989). If we assume that the luminosity
distribution of the bulge corresponds to the Gaussian components
rounder than $q'_j = 0.3$, we find that the bulge makes up $\sim 52$
per cent of the total luminosity of NGC~4342. The outer disk,
described by Gaussian components~9 and~10, makes up an additional
$46.5$ per cent, and the nuclear disk (modeled by Gaussian component
4) adds only about $1.5$ per cent to the total luminosity. There is no
reason that the mathematical components correspond to actual physical
components, but at least this gives an order-of-magnitude description
of the luminosities of the bulge and the two disk components. A more
accurate disk-bulge decomposition, which yields similar results, is
discussed in Scorza \& van den Bosch (1998).

Assuming that the density is built up from a sum of three-dimensional
Gaussians stratified on spheroids, one can, for any inclination angle
$i$, analytically calculate the density distribution from the MGE fit
to the intrinsic surface brightness. The mass density of such an MGE
model is given by
\begin{equation}
\label{mgedens}
\rho(R,z) = \Upsilon \sum_j I_j {\rm exp}\Biggl[ -{1\over 2 a_j^2}
\biggl( R^2 + {z^2 \over q_j^2} \biggr) \Biggr],
\end{equation}
where $\Upsilon$ is the mass-to-light ratio, and $I_j$, $a_j$ and
$q_i$ are related to $I'_j$, $a'_j$, $q'_j$ and $i$ (see EMB94).  The
potential that corresponds to this density distribution follows from
solving the Poisson equation. This yields
\begin{equation}
\label{mgepot}
\Phi(R,z) = -4 \pi G \Upsilon \sum_{j} a_j^2 q_j I_j
\int_{0}^{1} \exp\Biggl[ -{t^2 \over 2 a_j^2}  
\biggl( R^2 + {z^2 \over 1 - e_j^2 t^2} \biggr) \Biggr] 
{{\rm d}t \over \sqrt{1 - e_j^2 t^2}},
\end{equation}
where $e_j^2 = 1 - q_j^2$. 

The inclination angle is well constrained by the thinness of the
nuclear disk: $i > 83 \deg$ (Scorza \& van den Bosch 1998).
Throughout we assume that NGC~4342 is observed edge-on (i.e.,
$i=90\deg$).  Given the lower limit on the inclination angle of
$83\deg$, this assumption does not significantly influence the
conclusions presented in this paper.

\section{Jeans modeling}

\subsection{Formalism}

The three-integral modeling described in the next section requires
large amounts of CPU time. We therefore decided to first explore
parameter space of the models (i.e., mass-to-light ratio and mass of
the possible BH) by solving the Jeans equations and assuming that the 
phase--space distribution function depends only on the two classical integrals of
motion. The Jeans equations for hydrostatic equilibrium are moment
equations of the collisionless Boltzmann equation (see Binney \&
Tremaine 1987).  They relate the velocity dispersion tensor
${\widehat{\sigma}}^2$ and the streaming motion ${\bf v}$ to the
density $\rho$ and potential $\Phi$.  For an axisymmetric system with
distribution function $f(E,L_z)$ one always has $\sigma_R = \sigma_z$
and $\overline{v_R v_z} = 0$, and the Jeans equations in cylindrical
coordinates reduce to
\begin{equation}
\label{jeanseq}
{\partial (\rho \sigma_R^2) \over \partial R} + \rho \biggl(
{\sigma_R^2 - \overline{v_{\phi}^2} \over R} + {\partial \Phi \over
  \partial R} \biggr) = 0,
\end{equation}
\begin{equation}
\label{jeansequat}
{\partial(\rho \sigma_z^2) \over \partial z} + \rho {\partial \Phi \over
\partial z} = 0.
\end{equation}
Here $\; {\bar .} \;$ denotes the local average over velocities.  From
equation~(\ref{jeansequat}) and $\rho(R,z)$ and $\Phi(R,z)$, one can,
at every point $(R,z)$ in the meridional plane, calculate $\sigma_R^2$
($= \sigma_z^2$) by simple integration. By rewriting
equation~(\ref{jeanseq}), one can compute $\overline{v_{\phi}^2} =
{\overline{v_{\phi}}}^2 + \sigma_{\phi}^2$ without the need of
performing a numerical (ill-conditioned) derivative (see e.g. Hunter
1977; Simien, Pellet \& Monnet 1979; Binney, Davies \& Illingworth
1990).

The expressions for $\sigma_R^2$ and $\overline{v_{\phi}^2}$ for the
density distribution of equation (\ref{mgedens}) are given in EMB94
(their equations~[42] and~[44]).  The Jeans equations do not prescribe
how $\overline{v_{\phi}^2}$ splits in streaming motion
$\overline{v_{\phi}}$ and azimuthal velocity dispersion
$\sigma_{\phi}$. We follow the approach introduced by Satoh (1980),
and write
\begin{equation}
\label{satoheq}
\overline{v_{\phi}} = k \sqrt{\overline{v_{\phi}^2} - \sigma_R^2},
\end{equation}
such that we can control the anisotropy $\sigma_{\phi}/\sigma_R$ by
means of the free parameter $k$. For $k=1$ the model is fully
isotropic with $\sigma_{\phi} = \sigma_R = \sigma_z$. Once $k$ has
been fixed one can project the luminosity-weighted dynamical
quantities on the plane of the sky $(x',y')$ to yield the projected
rotation velocities
\begin{equation}
\label{vroteq}
V_{\rm rot}(x',y') = {-1\over S(x',y')} \int_{-\infty}^{\infty} \nu \,
\overline{v_{\phi}} \sin i \cos\phi \; {\rm d}z',
\end{equation}
and the rms velocities 
\begin{equation}
\label{vrmseq}
V_{\rm rms}^2(x',y') = {1\over S(x',y')} \int_{-\infty}^{\infty} \nu
\Bigl( \sigma_z^2 \cos^2 i + \sigma_R^2 \sin^2\phi \sin^2i +
(\sigma_{\phi}^2 + \overline{v_{\phi}}^2) \cos^2\phi \sin^2i \Bigr) \;
{\rm d}z'.
\end{equation}
Here $\nu = \rho/\Upsilon$ is the luminosity density and $S(x',y')$ is
the projected surface brightness at position $(x',y')$. $V_{\rm rot}$
and $V_{\rm rms}^2$ are the true first and second order moments of the
line-of-sight velocity distribution. The projected velocity dispersion
$\sigma_{p}(x',y')$ is simply derived from
\begin{equation}
\label{projsig}
\sigma_{p}(x',y') = \sqrt{V_{\rm rms}^2(x',y')-V_{\rm rot}^2(x',y')}.
\end{equation}
It is straightforward to include a BH in such a model, by simply
adding $GM_{\rm BH}/\sqrt{R^2+z^2}$ to the stellar potential
(\ref{mgepot}).

\subsection{Application to NGC 4342}

We use the Jeans equations to calculate the predicted rotation
velocities and velocity dispersions for the luminosity distribution of
NGC 4342. We assume that the stellar mass-to-light ratio $\Upsilon$
and the anisotropy parameter $k$ are constant throughout the
galaxy. We calculate $V_{\rm rot}$ and $V_{\rm rms}^2$ (using
equations [\ref{vroteq}] and [\ref{vrmseq}]) on a two-dimensional grid
on the sky. The grid is logarithmic in $r$ (in order to properly
sample the strong gradients near the center), and linearly sampled in
$\theta$. Once $V_{\rm rot}$ and $V_{\rm rms}^2$ are tabulated, we
convolve them with the PSF of the observations, weighted by the
surface brightness. After pixel binning, taking the proper slit width
into account, these are compared to the observations.

The $V$ and $\sigma$ determined from the GH fitting to the WHT VPs
cannot be compared directly to $V_{\rm rot}$ and $\sigma_p$ derived
from the modeling discussed above: the latter ones correspond to the
true moments of first and second order of the VPs, whereas the former
ones correspond to the best-fitting Gaussian. We therefore
recalculated the VP from $V$, $\sigma$, $h_3$ and $h_4$, from which we
then estimate the first and second order moments for direct comparison
with the Jeans models.

The results are shown in Figure~\ref{fig:jeansmod}, where we plot
$V_{\rm rot}$, $V_{\rm rms}$ and $\sigma_p$ of the VPs along the major
axis for both the WHT and the FOS data. Also plotted are predictions
for four models, that only differ in the mass of the central BH (0, 3,
5 and $10 \times 10^8 \Msun$). All models have $i=90\deg$, $\Upsilon_I
= 6.2 \Msun/\Lsun$ and $k = 1$ (i.e., all models are fully isotropic).
For this value of $k$, we obtain the best fit to the observed velocity
dispersion outside $\sim 2''$. However, $V_{\rm rot}$ is not very well
fitted: the wiggles in the rotation curve are not reproduced by the
model.  One can alter $k$, as function of radius, such that we fit
these wiggles, but at the cost of introducing them in the velocity
dispersion profile.  This is due to the poor fit of the Jeans models
to the rms velocities (see lower-left panel of
Figure~\ref{fig:jeansmod}).  $V_{\rm rms}^2$ depends only on the sum
of $\sigma_{\phi}^2$ and $\overline{v_{\phi}}^2$ (see
equation~[\ref{vrmseq}]) and is therefore independent of $k$.
Consequently, the Jeans models cannot simultaneously fit $V_{\rm rot}$
and $\sigma_p$ along the major axis. This suggests that the assumption
made, i.e., $f=f(E,L_z)$, is wrong, and that three-integral models are
required.

The Jeans models without a central BH clearly underpredict the central
velocity dispersion (for both the WHT and the FOS measurements), as
well as the central rotation gradient measured with the FOS. Models
with a massive BH provide a much better fit. The actual mass of the BH
depends on the data set used to constrain the model: the WHT data
suggests a BH mass of $\sim 3 \times 10^8 \Msun$, whereas the FOS data
are best fitted with $M_{\rm BH} \approx 6 \times 10^8 \Msun$.  So
although the two-integral Jeans modeling cannot fit all the observed
kinematics, it does suggest that a BH of a few times $10^8\Msun$ may
be present in the center of NGC~4342.

In Figure~\ref{fig:jeanscorr} we plot the ratio of the observed
rotation velocity over the rotation velocity of the best-fitting
isotropic Jeans model, $V_{\rm obs}/V_{\rm mod}$ (along the major
axis), versus the local observed ellipticity (in $I$-band) of
NGC~4342. There is a clear correlation in the sense that the Jeans
models underpredict the rotation velocity in the strongly flattened
region, and overpredict $V_{\rm rot}$ in the less flattened
region. The ellipticity is a measure of the local disk-to-bulge ratio,
and this therefore suggests that disk and bulge have different
velocity anisotropies. Since $V_{\rm rot}$ scales with
$\sqrt{\Upsilon}$, another possibility may be that the disk and bulge
are made up of different stellar populations (whose mass-to-light
ratios are different by almost a factor two). However, the separate
components (bulge, nuclear disk and outer disk) do not stand out as
separate entities in either the $U-V$ or the $V-I$ color images (see
BJM98), rendering this explanation improbable.

\section{Three-integral modeling}

In order to investigate the presence of a $\sim 3$--$6 \times
10^8\Msun$ BH as suggested by the Jeans models, we now construct fully
general, axisymmetric models of NGC~4342. We use an extension of
Schwarzschild's (1979) orbit-superposition method (see de Zeeuw 1997),
developed by Rix \etal (1997, hereafter R97) and Cretton \etal (1998,
hereafter C98).  The main method, but for spherical systems, is
outlined in R97.  The application to axisymmetric systems is discussed
in C98. Here we briefly outline the method, and we refer the
interested reader to R97 and C98 for details and tests of this
modeling technique.

\subsection{The method}

The first step of the method is to integrate orbits in the combined
potential $\Phi_{\rm stars} + \Phi_{\rm BH}$. Each orbit is then
projected onto the space of observables, taking convolution with the
PSF and pixel binning into account. Finally, a non-negative
least-squares algorithm is used to determine the distribution of orbit
weights that best fits the observational data (taking the
observational errors into account), while also reproducing the
luminous density distribution of the model.

Throughout we limit ourselves to models with an inclination angle $i =
90\deg$, and we assume that the stellar population has a constant
mass-to-light ratio. Therefore, each model is characterized by only
two free parameters: the mass-to-light ratio, $\Upsilon_I$, and the
mass of the black hole, $M_{\rm BH}$. Our aim is to find the set
$(M_{\rm BH},\Upsilon_I)$ that best fits the available constraints
(surface brightness and velocity profiles).

\subsection{The orbit library}

The motion of a star in an axisymmetric potential, for which $E$ and
$L_z = R v_{\phi}$ are conserved, can be reduced to motion in the
meridional $(R,z)$-plane, in the effective potential $\Phi_e =
\Phi(R,z) + L_z^2/(2 R^2)$. The orbit is constrained within a region
bounded by the zero velocity curve (ZVC) defined through
$E=\Phi_e(R,z)$.

Each orbit in an axisymmetric potential $\Phi(R,z)$ admits two
integrals of motion: energy $E=\Phi(R,z) + {1\over 2} v^2$ and
vertical angular momentum $L_z = R v_{\phi}$.  Regular orbits admit
one additional integral, $I_3$, which in general is not known
analytically. Such an orbit is confined to a sub-space inside the
ZVC. We only found a very small fraction of our orbit catalog to be
irregular.

The orbit library has to be set up such that one properly samples the
full extent of phase space. The sampling has to be sufficiently dense
to suppress discreteness noise, but is limited by the amount of
available CPU time. After some testing we have chosen to calculate
orbits on a $20\times 20 \times 7$ $(E,L_z,I_3)$-grid.  Each energy is
uniquely defined by a circular radius $R_c$ according to
\begin{equation}
\label{enereq}
E = {1\over 2} R_c {\partial \Phi \over \partial R}\biggl\vert_{R=R_c} +
\Phi(R_c,0).
\end{equation}
The 20 energies are sampled logarithmically between $R_c(E) = 0.01''$
and $R_c(E) = 60''$. These values were chosen such as to encompass the
major fraction of the total mass of the galaxy. The mass inside $R_c =
0.01''$ is only a fraction of $3.25 \times 10^{-5}$ of the total mass,
whereas a fraction of $\sim 2.85 \times 10^{-6}$ of the mass is
located outside $R_c = 60''$. For each energy, we calculate the
maximum vertical angular momentum
\begin{equation}
\label{lzeq}
L_{z,{\rm max}}(E) = \sqrt{R_c^3 \; 
\biggl({\partial\Phi \over \partial R}\biggr)_{(R_c,0)}},
\end{equation}
which corresponds to the circular orbit with energy $E$, and sample
$\vert \eta \vert = \vert L_z \vert /L_{z,{\rm max}}$ on a linear grid
of 10 values between 0.01 and 0.99. Hence, the purely circular and
radial orbits are presumed to be represented by their closest
neighbors on the grid, but are not explicitly calculated.  At each
value of $\vert \eta \vert$, only the orbit with positive angular
momentum has to be integrated, since its counterpart ($L_z = -\eta
L_{z,{\rm max}}$) is simply a mirror reflection around zero velocity.
Since the third integral $I_3$ can generally not be expressed
explicitly in terms of the phase-space coordinates, we use the method
suggested by Levison \& Richstone (1985), and take the starting point
on the ZVC as a numerical representation of the third integral (see
also C98). For that purpose we first calculate, for each
$(E,L_z)$-pair, the locus $(R_{\rm tt}, z_{\rm tt})$ on the ZVC of the
`thin tube orbit'.  For a certain value of $E$ and $L_z$, this is the
only orbit that touches the ZVC at only one value of $R$. All other
regular orbits touch the ZVC at two different values of $R$. We sample
$I_3$ by linearly sampling the $R$-coordinate of the orbit's starting
point on the ZVC ($R_{\rm zvc}$) between $R_{\rm tt}$ and $R_{\rm
max}$, where $R_{\rm max}$ is the maximum extent of the ZVC in the
equatorial plane.

Since the calculation of the potential (equation~[\ref{mgepot}]) and
the forces all require the evaluation of a numerical quadrature, we
have calculated them on a $4000 \times 300$ $(R,\theta)$-grid in the
meridional plane. The grid is sampled linearly in $\theta$ between $0$
and $\pi/2$, and logarithmically in $R$ between $10^{-4}$ and $10^{3}$
arcseconds.  Each orbit is integrated for 200 radial periods, using
linear interpolation between grid points to evaluate the potential and
forces. On average, the energy conservation over 200 radial periods is
better than one part in $10^5$, justifying the interpolation scheme
adopted. In total $20 \times 10 \times 7 = 1400$ orbits are
integrated, resulting in a library of $2 \times 1400 = 2800$ orbits
(when doubling for the $-L_z$ orbits).

During the integration of each orbit we project its phase-space
coordinates onto the space of observables $(x',y',v_{\rm los})$,
where $(x',y')$ is the plane on the sky, and $v_{\rm los}$ is the
line-of-sight velocity. In the following we use $v$ as shorthand for
$v_{\rm los}$. We adopt a three-dimensional grid in the
$(x',y',v)$-space, i.e., our storage cube, in which we record the
fractional time the orbit spends in each of the cells (see R97 for
details). Once the orbit integration is finished, we convolve each of
the velocity slices $(x',y')$ with the PSF appropriate for the
observations.  Since we have kinematic constraints obtained with three
different instrumental setups and different PSFs (WHT major axis, WHT
minor axis, and FOS apertures) we use three separate
$(x',y',v)$-storage cubes each of which is convolved with the
respective PSF.  We use two cubes with $0.1'' \times 0.1''$
$(x',y')$-cells for the WHT major and minor axes. For the FOS-cube,
$0.05'' \times 0.05''$ cells are used to comply with the higher
spatial resolution of the HST. For all storage cubes we use 101
velocity bins of $30 \kms$. The final step is to calculate the
contribution of the orbit to each of the positions in the plane of the
sky where we have photometric and/or kinematic constraints. These
positions are in general extended areas (e.g., determined by the pixel
size of the CCD, the slit width and the pixel rebinning used to obtain
spectra of sufficient $S/N$). For each constraint position
$l=1,...,N_c$ we therefore sum the fractional times over the area of
$l$ (see R97 for details). This gives us in the end, for each orbit
$k$ and each constraint position $l$, the properly PSF-convolved
velocity profile `histogram' ${\rm VP}^k_{l,v}$, integrated over the
area of position $l$.  By using {\it fractional} times we ensure that
each orbit is normalized to unity.

\subsection{The observational constraints}

The final step of the orbit-superposition method is to find the set of
non-negative weights, $\gamma_k$, of each orbit that best matches the
kinematic constraints and reproduces the luminous mass density of the
model. Since we normalize each orbit to unity, the orbital weights
$\gamma_k$ measure the fraction of the total light of the galaxy that
resides on orbit $k$. We use the following sets of constraints:

The solution ${\bf \gamma}$ has to reproduce the luminosity density
$\rho(R,z)/\Upsilon$ (equation~[\ref{mgedens}]). We have subdivided
the first quadrant of the meridional plane in $20\times 7$
$(r,\theta)$-cells, with $r$ and $\theta$ the standard spherical
coordinates. The grid that encompasses those cells is binned linearly
in $\theta$, and logarithmically in $r$ between $0.01''$ and $60''$.
For each orbit $k$ we store the fractional time $t^k_n$ spent in cell
$n$. Since we integrate the orbits in the meridional plane, this is
similar to the fractional time the orbit spends in the
three-dimensional volume obtained by integrating the area of the cell
over $2\pi$ radians in the $\phi$-direction. Therefore we have
computed the total luminosity, $L_n$, of the MGE model inside each
volume $n$. We will refer to these constraints as the
``self-consistency constraints''.

From the WHT spectra we obtained sets of $(V_l,\sigma_l,h_{3,l},
h_{4,l})$ at 19 positions $l$ along the major axis, and 8 along the
minor axis. The quantities are the luminosity-weighted averages over
the areas of constraint positions $l=1,...,27$. For each of these 27
positions we have calculated the surface brightness $S_l$ integrated
over that area, and convolved with the appropriate PSF. The quantities
$S_l$, $V_l$, $\sigma_l$, $h_{3,l}$ and $h_{4,l}$ parameterize the
velocity profiles ${\cal L}_l^{\rm obs}(v)$ with
\begin{equation}
\label{sleq}
S_l = \int_{-\infty}^{\infty} {\cal L}_l^{\rm obs}(v) {\rm d}v.
\end{equation}
One can rewrite this parameterization in the form $(S_l, S_l h_{1,l},
S_l h_{2,l}, S_l h_{3,l}, S_l h_{4,l})$, with
\begin{equation}
\label{newsleq}
S_{l} h_{m,l} = 2 \sqrt{\pi} \int_{-\infty}^{\infty} {\cal
  L}_{l}^{\rm obs}(v) \alpha(w_l) H_m(w_l) {\rm d}v,
\end{equation}
where $m=1,...,4$, $\alpha$ is again the standard Gaussian (see
equation~[\ref{alphaweq}]), and
\begin{equation}
\label{wleq}
w_l = {v-V_l \over \sigma_l},
\end{equation}
with $V_l$ and $\sigma_l$ the measured rotation velocity and velocity
dispersion of the VP's best-fitting Gaussian at constraint position
$l$. Note that with this definition, $h_{1,l}$ and $h_{2,l}$ are zero
for all constraint positions.  This parameterization has the advantage
that the orbit-superposition problem for the orbit weights is linear
(see R97).

In addition, we obtained $(V_l,\sigma_l)$ at 7 FOS aperture-positions.
Again we have calculated, for each of these positions $l$, the
PSF-convolved, aperture-integrated surface brightness $S_l$. The PSF
used and the aperture diameter adopted are described in BJM98.  As for
the WHT spectra, we use the parameterization ($S_{l}$, $S_{l} h_{m,l}$
with $m=1,2$) as constraints rather than ($S_{l},V_l,\sigma_l)$.

In total we thus have 140 self-consistency constraints $L_n$, 34
constraints on projected surface brightness $S_l$, and 122 kinematic
constraints $S_l h_{m,l}$.

\subsection{Non-negative least squares fitting}

At each constraint position $l$ we have a measured velocity profile
${\cal L}_l^{\rm obs}(v)$, which we parameterized by $(S_l,S_l
h_{m,l})$. For each of the observational constraints, we have the
measurement errors $\Delta S_l, \Delta V_l, \Delta \sigma_l, \Delta
h_{3,l}$ and $ \Delta h_{4,l}$ from which we can calculate the errors
$\Delta(S_l h_{m,l})$. As described in Section~5.2 we determined, for
each orbit $k$ and each position $l$, the velocity profile ${\rm
VP}^k_{l,v}$ at velocity $v$. We parameterize each orbital VP by
$(S^k_l, S^k_l h^k_{m,l})$ with $m=1,...,4$ using
equations~(\ref{sleq}) and~(\ref{newsleq}) and with ${\cal L}_{l}^{\rm
obs}(v)$ replaced by ${\rm VP}^k_{l,v}$, and by changing the
integration to a summation over all velocity bins. The orbit weights
$\gamma_k$ ($k=1,...,N_{\rm orbits}$) that result in the best fit to
the observations can be determined by minimizing
\begin{equation}
\label{chi2eq}
\chi^2_{\rm obs} = \sum_l \Biggl( {S_l - \sum_k \gamma_k S^k_l \over
  \Delta S_l} \Biggr)^2 + \sum_{m=1}^4 \sum_l \Biggl( {S_l h_{m,l} -
  \sum_k \gamma_k S^k_l h^k_{m,l} \over \Delta (S_l h_{m,l})}
\Biggr)^2.
\end{equation}
In addition to minimizing $\chi^2_{\rm obs}$, we also want the
solution to match the luminosity density in the meridional plane,
i.e., we also want to minimize
\begin{equation}
\label{newchi2eq}
\chi^2_{\rm sc} = \sum_n \Biggl( {L_n - \sum_k \gamma_k t^k_n \over 
\Delta L_n} \Biggr)^2, 
\end{equation}
where $\Delta L_n$ sets the accuracy for reproducing the luminosity
density in the meridional plane. Throughout, we set $\Delta S_l = 0.01
S_l$ and $\Delta L_l = 0.01 L_l$, such that we aim for an accuracy of
one per cent in reproducing both the projected PSF-convolved surface
brightness and the luminosity density in the meridional plane. It is
in principle sufficient to fit only the luminous density in the
meridional plane: the surface brightness should be fitted
automatically.  However, because of discretization this is in practice
not necessarily so, and we thus include the surface brightness at the
constraint positions $l$ as separate constraints.

Minimizing $\chi^2 \equiv \chi^2_{\rm obs} + \chi^2_{\rm sc}$ is a
least-squares problem for a matrix equation (see R97 for a detailed
description of the matrices involved).  It has to be solved under the
physical constraint that $\gamma_k \geq 0$.  Following Pfenniger
(1984), R97 and C98, we use the Non-Negative Least Squares (NNLS)
algorithm by Lawson \& Hanson (1974) to solve for the orbit weights.

\subsection{Regularization}

The NNLS matrix equation solved is numerically rather
`ill-conditioned' giving rise to a distribution function with strong
oscillatory behavior. Such DFs are unphysical (e.g., Lynden-Bell
1967; Spergel \& Hernquist 1992; Merritt 1993).  Smoothing in the
solution space can be achieved via regularization (e.g., Merritt 1996
and references therein).  We follow the scheme used by Zhao (1996),
which is based on a minimization, up to a certain degree, of the
differences in weights between neighboring orbits.  The technique is
described in R97, C98, and van der Marel \etal 1998, and we refer the
interested reader to those papers for details.  The extra
regularization constraints result in a less good fit to the data. The
amount of smoothing is set such that the regularized model is still
compatible with the data in a statistical sense, i.e., such that
$\Delta \chi^2 = \chi^2 - \chi^2_{\rm min} = 1$, with $\chi^2_{\rm
min}$ the value obtained without regularization. Unless mentioned
otherwise, we discuss results in which no regularization has been
adopted.

\section{Shortcoming of the Gauss-Hermite parameterization}

Although the modeling technique outlined in the previous sections
works well for dynamically hot galaxies (e.g. M32, see van der Marel
\etal 1998), some problems arise when trying to apply it to
dynamically cold systems such as NGC~4342.  This is due to the fact
that we do not include the zeroth-order moment $h_0$ in the fit. This
quantity measures the normalization of the best-fitting Gaussian to
the normalized VP and is observationally inaccessible: it is directly
proportional to the unknown difference in line strength between the
galaxy spectrum and the template spectrum used to analyze. (In
practice one uses the assumption $h_0=1$ to estimate the line strength
from the observations.)  As we now show, excluding $h_0$ from the GH
series and expanding the orbital VPs around the observed VPs up to
fourth order only, can lead to artificial counter-rotation in models
of cold systems.

Assume we have an observed velocity profile ${\rm VP}_{\rm obs}$, at a
certain position on the sky, that is perfectly Gaussian. The GH
moments $h_m$ with $m>0$ of such a ${\rm VP}_{\rm obs}$ will all be
zero. In the method described above, we derive the orbital $h^k_m$
($m=1,...,4$) from equation~(\ref{newsleq}) in which the observed $V$
and $\sigma$ enter in the weighting function $\alpha(w) H_m(w)$. In
the NNLS algorithm we solve for the orbit weights by minimizing the
difference between the GH moments of ${\rm VP}_{\rm obs}$ and ${\rm
VP}_{\rm orb}$. In principle, the differences between the $S_l
h_{m,l}$ and $\sum_k \gamma_k S^k_l h^k_{m,l}$ are minimized, but for
simplicity we illustrate the problem with a one-orbit model.

In the ideal case, an orbit whose VP deviates more strongly from the
observed VP will be assigned a smaller weight. However, this is not
always the case with the VP parameterization described in Section~5.3.
To illustrate this we calculated the GH moments $h_m$ ($m=0,...,2$) of
a Gaussian expanded around another Gaussian, both with the same
dispersion $\sigma$. In Figure~\ref{fig:gh} we plot the resulting
moments $h_m$ as function of the velocity difference $\Delta V$
between the two Gaussians (in units of $\sigma$). For $\Delta V = 0$,
$h_0 = 1$ and all higher-order moments are zero (i.e., the two
Gaussians are identical).  In the regime $\vert \Delta V \vert /
\sigma \lesssim 2$ the higher-order moments increase with increasing
velocity-difference. For larger values of $\vert \Delta V \vert /
\sigma$ they start to decrease again to reach approximately zero for
$\vert \Delta V \vert / \sigma \gtrsim 5$.  On the contrary, $h_0$ equals
unity when the VPs are identical, and decreases monotonically for
increasing $\vert \Delta V \vert / \sigma$.

In NGC~4342, which is a dynamically cold system, the major-axis
kinematics reach $V_{\rm obs}/\sigma_{\rm obs} > 2.5$ at the outside
(see Figure~\ref{fig:velsig}).  An orbital VP with approximately the
same $V$ and $\sigma$ as an observed VP with $V/\sigma > 2.5$ will
have its $h_1$ and $h_2$ close to zero. Consequently, it will likely
be given a non-zero weight. The same orbit, but with opposite sense of
rotation (i.e., with reversed vertical angular momentum), will also
have $h_1$ and $h_2$ close to zero, since $\vert \Delta V \vert /
\sigma > 5$.  In other words, the $+L_z$ and $-L_z$ orbits have the
same $S$, $h_1$ and $h_2$ and are therefore indistinguishable for the
fitting algorithm. The $+L_z$ and $-L_z$ orbits have opposite values
of $h_3$, but if $\vert h_3 \vert$ is small, the difference between
the two orbits, in terms of the VP parameterization used, remains
small.  If it would have been possible to include $h_0$ in the fit,
then the $+L_z$ would have $h_0$ close to unity, and the $-L_z$ orbit
would have $h_0 \approx 0$ (see Figure~\ref{fig:gh}): the two VPs {\it
would} have been easily distinguishable. In addition, if the
observational data had been of sufficient $S/N$ to derive GH moments
up to very high order, it would also have been possible to distinguish
the VPs of the $+L_z$ and $-L_z$ orbits. In practice, however, one
requires unrealistic high $S/N$ spectra to be able to measure these
moments.

We can thus expect that our solutions will have significant amounts of
counter-rotation at radii where $V_{\rm obs}/\sigma_{\rm obs} \gtrsim 2$.
We indeed found solutions in which the reconstructed VPs at large
radii along the major axis have two peaks at positive and negative
$v_{\rm los}$ (see Section~7.1). Van der Marel \etal (1998) applied
the same modeling technique to the dynamically hot system M32, which
has $V/\sigma < 1.0$ everywhere, and therefore did not encounter this
problem.

To solve the problem outlined above, we use a `modified approach' in
which we add some extra constraints to the models: we exclude
counter-rotating orbits whose circular radius $R_c$ is larger than a
limiting radius $R_{\rm lim}$. We have chosen $R_{\rm lim} = 4.5''$,
since for larger radii $V_{\rm obs}/\sigma_{\rm obs} > 1.5$ (see
Figure~\ref{fig:velsig}) and the VP parameterization used becomes poor
in the light of the problem discussed above. As we show in
Section~7.1, this solves the problem of the artificial
counter-rotation at large radii along the major axis, but it has the
disadvantage that our models are not fully general any more: we have
imposed some a priori constraints on the amount of counter-rotating
orbits in NGC~4342.  On the other hand, the actual observations do not
reveal any counter rotation at large radii. Critically, one may argue
that the VP analysis of the spectra is not suitable to detect such
counter rotation, since we do not go to sufficient high order in the
GH expansion. We have therefore also analyzed our spectra with the
unresolved Gaussian decomposition (UGD) method (Kuijken \& Merrifield
1993), which is not hampered by this limitation (e.g., Merrifield \&
Kuijken 1994), and found good agreement with the results of the GH
parameterization. This suggests that indeed NGC~4342 has a negligible
amount of counter-rotating stars in its outer disk.  The additional
constraint imposed on the models, when using the modified approach, is
thus justified observationally.

\section{Results and discussion}

\subsection{The black hole and mass-to-light ratio}

Based on the BH mass and mass-to-light ratio suggested by the Jeans
modeling (Section~4.1) we construct three-integral models with $M_{\rm
BH}/\Upsilon_I$ in the range $0 - 1.5 \times 10^8 \Msun$ and $4 \leq
\Upsilon_I \leq 9$. For each value of $M_{\rm BH}/\Upsilon_I$ only one
orbit-library has to be constructed: a change in mass-to-light ratio
is equivalent to a scaling of the model velocities proportional to
$\sqrt{\Upsilon}$. We therefore calculate 10 different
orbit libraries, all with $\Upsilon_I = 1$, that differ only in the
mass of the central BH. If one scales to another value of
$\Upsilon_I$, the mass of the BH changes accordingly to $\Upsilon_I \,
M_{\rm BH}$. We sample the mass-to-light ratio at 16 different values
in the interval $\Upsilon_I \in [4,9]$.

For each of the in total 160 different ($M_{\rm
  BH},\Upsilon_I$)--models we minimize $\chi^2 = \chi^2_{\rm obs} +
\chi^2_{\rm sc}$ (equations~[\ref{chi2eq}] and~[\ref{newchi2eq}]).  We
use $\chi^2$ statistics to compare different $(M_{\rm
  BH},\Upsilon_I)$--models in a proper statistical way.  We determine
the measure $\Delta \chi^2 \equiv \chi^2 - \chi^2_{\rm min}$, where
$\chi^2_{\rm min}$ is the overall lowest $\chi^2$.  Under the
assumptions that the errors are normally distributed and that there
are no numerical errors in the model, one can assign confidence levels
to the measure $\Delta \chi^2$. The exact level of confidence depends
on the number of degrees of freedom in the models (Press \etal 1992),
in our case two: $M_{\rm BH}$ and $\Upsilon_I$.

The resulting $\chi^2$--plots are shown in Figure~\ref{fig:chicont}.
The first three contours show the formal $68.3$, $95.4$ and $99.73$
per cent confidence levels (the latter one is plotted with a thick
contour).  The solid dots correspond to actual model calculations.
Bi-cubic spline interpolation is used to calculate $\chi^2$ at
intermediate points. Four different plots are shown, labeled $a$ to
$d$. Plot $a$ (upper-left panel) shows the $\chi^2$ plot of the fits,
when using only the $V$ and $\sigma$ measurements of the WHT spectra
as constraints. Clearly, a large range of parameter space gives
equally good fits. There is at best only a marginal indication that
models with a BH fit the ground-based rotation velocities and velocity
dispersions better than without a BH. Plot $b$ (upper-right panel) is
similar to plot $a$, except that we have now included the $h_3$ and
$h_4$ measurements of the WHT spectra as constraints on the models.
Two changes are evident. First, the allowed range of $\Upsilon_I$ at
given BH mass is much smaller. Secondly, BHs more massive than $\sim 6
\times 10^8 \Msun$ can be ruled out at the $99.73$ per cent confidence
level. Plot $c$ (lower-left panel) shows the resulting $\chi^2$-plot
when all constraints, including the HST measurements are taken into
account. These high spatial-resolution measurements allow us to rule
out models without a BH at a confidence level better than $99.73$ per
cent. Finally, in plot $d$ (lower-right panel) we show the results
when using the modified approach on the entire set of constraints. As
can be seen, the exclusion of counter-rotating orbits with circular
radii beyond $4.5''$ puts some further limits on the allowed range of
acceptable models, consequently contracting slightly the $99.73$ per
cent confidence region.

Although the solutions we find with the standard approach can result
in models with significant counter-rotation (see Section~6), one can
still meaningfully use the $\chi^2$ statistics to put confidence
levels on $M_{\rm BH}$ and $\Upsilon_I$. The main requirement is that
orbital VPs and observed VPs are parameterized in exactly the same
way, which is the case. The counter rotation that we find is simply a
consequence of our particular VP parameterization. Although the
modified approach results in a more strict solution-space, and in
principle is based on additional constraints that are observationally
justified, we will nevertheless consider the $\chi^2$ surface of
Figure~\ref{fig:chicont}$c$ as the main result.  We merely present the
results of the modified approach to show that the counter-rotation in
the best-fitting models has no significant influence on the BH mass
and mass-to-light ratio of our best-fitting model.  Furthermore,
Figure~\ref{fig:chicont}$c$ results in more conservative estimates of
the errors on BH mass and mass-to-light ratio.

The labeled asterisks in panel~$c$ indicate three models that will be
discussed in more detail below. Model~B provides the best overall fit
and has $M_{\rm BH} = 3.6 \times 10^8 \Msun$ and $\Upsilon_I = 6.25
\Msun/\Lsun$.  From the $\chi^2$ statistics we derive $M_{\rm BH} =
3.0^{+1.7}_{-1.0} \times 10^8 \Msun$ and $\Upsilon_I =
6.3^{+0.5}_{-0.4} \Msun/\Lsun$. The errors are the formal 68.3 per
cent confidence levels. For comparison, the modified approach yields
$M_{\rm BH} = 3.0^{+0.9}_{-0.8} \times 10^8 \Msun$ and $\Upsilon_I =
6.5^{+0.3}_{-0.3} \Msun/\Lsun$, which is consistent with and slightly
more strict than the results obtained with the standard method.

Van der Marel \etal (1998) used the same three-integral modeling
technique and statistical means to infer that M32 harbors a massive
BH of $(3.4 \pm 0.7) \times 10^6 \Msun$. To facilitate direct
comparison, the $\chi^2$ surfaces presented here are plotted with the
same contour levels as in van der Marel \etal (1997, 1998). The $\chi^2$
contours of Figure~\ref{fig:chicont} correspond to the goodness-of-fit
to {\it all} constraints.  Van der Marel \etal plotted contours of
$\chi^2$ of the fit to the kinematic constraints only (i.e., the right
term in equation~[\ref{chi2eq}]), but noted that contour plots of the
total $\chi^2$ look similar. We have computed the surfaces of the
$\chi^2$ that only measures the fit to the kinematic constraints and
indeed found very similar contours: all models can fit the surface
brightness and luminosity density to better than one per cent as
requested.
 
In order to compare the kinematical predictions of the models with the
actual observations, we compute the velocity profile of the model at
each constraint position $l$, ${\rm VP}_{l,v}$, as the weighted sum of
all ${\rm VP}^k_{l,v}$, using the solution $\gamma$ for the orbit
weights.  For these VPs we then compute $V_l$ and $\sigma_l$ of the
best-fitting Gaussian, as well as $S_l$ and the GH moments $h_{3,l}$
and $h_{4,l}$, all of which can be directly compared to the
observations. The kinematical predictions of models A, B and C are
plotted in Figure~\ref{fig:modelfit} together with the observations.
Although we did not measure the $h_3$ and $h_4$ coefficients for the
FOS spectra, we plot the model predictions for these quantities.
All three models provide equally good fits to the projected surface
brightness and the meridional luminosity density (not plotted). The
main difference between models~A and~B is their fit to the central HST
velocity dispersion.  The model without BH (model A) underpredicts the
observed dispersion by $\sim 130\kms$ ($\sim 4.0 \Delta \sigma$). The
rapid central rotation in NGC~4342, as measured with the FOS is
remarkably well fit even without central BH.  The main difference
between models~B and~C is their fit to the central WHT velocity
dispersion along the minor axis, and the fit to the $h_3$ measurements
along the major axis. Although model~C fits the central HST velocity
dispersion even better than model~B, the fit to the higher-order GH
coefficients is worse to such an extent that this model can be ruled
out to better than 99.73 per cent confidence (cf.~panels $a$ and $b$
of Figure~\ref{fig:chicont}).

The large differences between the predictions of model~C at the
outermost point along the major axis (WHT measurements) and the actual
observed values is due to the problem with the counter-rotation. This
is illustrated in Figure~\ref{fig:doublepeak}, where we plot VPs at
four different radii along the major axis. Solid dots represent the
`observed' velocity profiles, reconstructed from the measured
GH-parameters (assuming that all GH coefficients of order five and
higher are zero). The solid lines represent the reconstructed VPs of
model C, and the dashed lines correspond to the Gauss-Hermite series
fitted to these model VPs. The upper four panels show the VPs of model
C.  The lower panels show the VPs for a model with the same BH mass
and mass-to-light ratio as model C, but for which the modified
approach is used. For $R=0.0''$ and $R=4.83''$ the model VPs of the
two different approaches are almost identical. However, for the larger
radii, the VPs of model C clearly reveal large amounts of counter
rotation. The VP of the upper-right panel has such a large
counter-rotating component, that the best-fitting Gaussian no longer
corresponds to the peak at positive velocities: it is very broad and
centered around $V \approx 0 \kms$.  This causes the large difference
between the model predictions and the observed values at the outermost
WHT point in Figure~\ref{fig:modelfit}. Although the reconstructed $V$
and $\sigma$ may differ strongly from the observed values, the $h_1$
and $h_2$ values are very similar. Since these are the values that
enter into the computation of $\chi^2$, this discrepancy does not
affect the $\chi^2$ statistics.
 
It is interesting that the central velocity dispersions, as measured
with the WHT, can be fit without the requirement of a central BH.
This contradicts the conclusions reached from the Jeans modeling,
which suggests, on the basis of the WHT measurements alone, that a BH
of $\sim 3 \times 10^8 \Msun$ is required. The three-integral
modeling, however, shows that only the HST measurements are of
sufficient spatial resolution to discriminate strongly between models
with and without a massive BH (cf.~panels $b$ and $c$ of
Figure~\ref{fig:chicont}).  Clearly, without these high
spatial-resolution kinematics, the case for a BH is only marginal: the
$\chi^2$ plot in panel $b$ of Figure~\ref{fig:chicont} suggests a BH
mass of $\sim 2.0 \times 10^8 \Msun$, but cannot rule out models
without a BH to a significant confidence level.

\subsection{Alternatives to a black hole}

\subsubsection{A dense cluster}

Although the dynamical evidence for the presence of a MDO of a few
$10^8 \Msun$ in NGC~4342 is compelling, it does not automatically
imply evidence for a BH.  Alternatives to a point mass, such as a
cluster of brown dwarfs or stellar remnants are not ruled out by the
modeling presented above.  Any of these alternatives is only viable if
its lifetime is not significantly smaller than the age of the galaxy
(typically $\sim 10$ Gyr). There are three main processes that
determine the evolution of a dense cluster: (i) core collapse, (ii)
evaporation due to weak gravitational scattering, and (iii) physical
collisions between the objects comprising the cluster. The latter of
these processes is likely to ultimately lead to the formation of a
single dense object, probably a massive BH. After a timescale
$\tau_{\rm coll}$, each object has physically collided with another
(in a statistical sense). $\tau_{\rm coll}$ strongly depends on the
mass and density of the cluster, as well as on the mass and size of
the constituents (see Maoz 1997 and references therein). The
characteristic timescale for evaporation of a cluster which consists
of equal mass constituents is $\tau_{\rm evap} \approx 300 \;
\tau_{\rm relax}$ (Spitzer \& Thuan 1972), whereas typically after
$\tau_{\rm cc} \approx 16 \tau_{\rm relax}$, a Plummer sphere of equal
mass objects undergoes core collapse (Cohn 1980). Here $\tau_{\rm
  relax}$ is the median relaxation time (see Spitzer \& Hart 1971).
By choosing a Plummer model for the dark cluster we are conservative,
in that more concentrated clusters have shorter collapse times, and
are thus less likely as alternatives for a massive BH.
 
In order to constrain the size of a dark cluster, we construct models
in which we replace the point-mass potential of the BH by a Plummer
potential with a scale length $\epsilon$. We consider model B for
which we replace the $3.6 \times 10^8 \Msun$ BH by a Plummer potential
with the same mass, but with different values of $\epsilon$.  The
stellar mass-to-light ratio is kept constant at 6.25.
Figure~\ref{fig:chieps} shows the resulting $\chi^2$ as function of
$\epsilon$. The best fit is obtained for $\epsilon = 0.0$ (model B),
and the fit deteriorates with increasing scalelength of the Plummer
potential. The dotted line indicates the formal 99.73 per cent
confidence level, and at this level of confidence we can rule out dark
clusters with $\epsilon > 0.07''$. This upper limit on the scale
length of the Plummer sphere corresponds to 5.1 pc at the assumed
distance of 15 Mpc, implying a central density of the cluster $> 6.7
\times 10^5 \Msun {\rm pc}^{-3}$.

We calculate the relaxation timescale $\tau_{\rm relax}$, and the
collision timescale $\tau_{\rm coll}$ in which we adopt the
mass-radius relation for {\it non-collapsed} objects used by Goodman
\& Lee (1989).  In Figure~\ref{fig:lifetime} we plot the
characteristic timescales for core-collapse and collisional
destruction of our dark cluster, as function of the mass $m$ of the
cluster's constituents. As can be seen, the timescales for core
collapse ($\tau_{\rm cc} \approx 7 \times 10^{11} (m/\Msun)^{-1}$ yr)
and evaporation ($\tau_{\rm evap} \sim 20 \tau_{\rm cc}$) for the
inferred dark cluster in NGC~4342 exceed the Hubble time for $m
\lesssim$ few $\Msun$, and neither of these processes thus allows us
to rule out a dark cluster as an alternative to a BH.  The collision
timescale $\tau_{\rm coll}$ is the most restricting, and we are close
to being able to rule out dark clusters of non-collapsed objects with
masses less than $\sim 0.001 \Msun$.  However, for clusters of brown
dwarfs with masses of $\sim 0.08 \Msun$, $\tau_{\rm coll}$ is still of
the order of $10^{12}$ yr, and such clusters can clearly not be ruled
out by current observations.  Clusters of {\it collapsed} stellar
remnants, such as white dwarfs or neutron stars, have collision
timescales that are even much larger than the value of $\tau_{\rm
  coll}$ plotted in Figure~\ref{fig:lifetime}.  This is due to the
much smaller collisional cross-sections of these collapsed objects.
In fact, the collapse time for a dark cluster can be made arbitrarily
long by giving its objects an arbitrarily small mass, and the problems
with collisions and mergers can be avoided by assuming the cluster to
be a collisionless gas of elementary particles. However, another
effect might allow one to rule out such a dark cluster: the trapping
of stars by the cluster due to dynamical friction. If the time scale
for this process is smaller than the age of the galaxy, enouh stars
will get trapped such that the cluster can no longer be considered
dark. Alternatively, these trapped stars can merge and form a massive
BH, thus spoiling the goal for which the cluster was introduced.  The
time scale for the trapping to occur is independent of the mass $m$ of
the cluster constituents, but does depend on the mass $m^*$ of the
stars being captured, and is given by $(m/m^*) \tau_{\rm cc}$ (as long
as $m^* \gg m$, see Quinlan 1996). For NGC~4342 we thus find that this
capture time scale is only smaller than $\sim 10^{10}$ yr for stars
with $m^* \gta 70 \Msun$. We can thus not use this trapping-mechanism
to put an appreciable constraint on the nature of a possible dark
cluster in NGC~4342.
 
In conclusion, it is clear that we are still a long way from being
able to rule out dark clusters as viable alternatives for a massive BH
in the center of NGC~4342.

\subsubsection{An end-on bar}

It has been argued that a bar observed end-on may mimic the presence
of a BH (Gerhard 1988). In particular, the nuclear disk in NGC~4342
may in fact be a (very thin) nuclear bar. However, there are several
reasons why this interpretation is unlikely. First of all,
axisymmetric models without BH fail to fit the central velocity
dispersion, as measured with the HST/FOS, by $\sim 130\kms$. It seems
unlikely that the elongated orbits in a bar potential can fix this.
Furthermore, there are no indications that (the center of) NGC~4342 is
triaxial. Although van den Bosch \& Emsellem (1998) have provided
evidence that NGC~4570, a galaxy with a double-disk morphology similar
to NGC~4342, has been shaped under the influence of a rapidly tumbling
bar-potential, none of the characteristics found in NGC~4570 that led
to this conclusion are apparent in NGC~4342.  In addition, no minor
axis rotation is found, and the small amount of isophote twist
observed (see BJM98) is limited to the outer region of the galaxy, and
is more likely to be associated with a small warping of the outer
disk, probably induced by the small companion at $\sim 30''$ SE.
Finally, the steep cusp of the bulge of NGC~4342 makes the bar
hypothesis unlikely, since the pressure support from this strongly
cusped bulge probably assures stability against bar formation (cf.~van
den Bosch \& de Zeeuw 1996).

\subsection{The dynamical structure of NGC~4342}

During the orbit integrations we store, in addition to the fractional
time spent by orbit $k$ in meridional cell $n$, $t^k_n$, the
time-weighted first and second order velocity moments averaged over cell
$n$. With the solution for the orbit weights $\gamma$ we can compute
the internal dynamical structure of the model averaged over each cell
$n$. These are given by 
\begin{equation}
\label{firstmom}
\langle v_a \rangle_n = {\sum_k \gamma_k \; t^k_n \; \langle v_a \rangle^k_n
\over \sum_k \gamma_k \; t^k_n},
\end{equation}
and 
\begin{equation}
\label{secondmom}
\langle v_a^2 \rangle_n^{} = {\sum_k \gamma_k \; t^k_n \; 
\langle v_a^2 \rangle^k_n \over \sum_k \gamma_k \; t^k_n}.
\end{equation}
Here we use the subscript $a$ to indicate either of the cylindrical
coordinates $R$, $\phi$ or $z$. The cell-averaged velocity dispersions
can be computed according to $\langle \sigma_a \rangle_n^{2} = \langle
v_a^2 \rangle_n^{} -\langle v_a \rangle^2_n$.

In order to suppress noise, we average $\langle v_a \rangle^{}_n$ and
$\langle v_a^2 \rangle^{}_n$ over cells $n$ that have the same radius
$r$ but different $\theta$ ($r$ and $\theta$ being the standard
spherical coordinates). Given the strongly flattened shape of NGC~4342
and its multi-component structure, we decided to split the galaxy in
two parts: we determine the internal dynamical structure in two cones;
one with half-opening angle of $30\deg$ centered around the equatorial
plane, including the nuclear and outer disks, and the other one with
half-opening angle of $60\deg$ centered around the minor axis,
representing the bulge of NGC~4342.

We investigate the dynamical structure of models A, B, and C using the
modified approach (see Section~6.1) and regularization as described in
Section~5.5 in order to suppress noise. As a check, we compare the
dynamical structures before and after regularization (with the amount
of smoothing chosen such that $\Delta \chi^2 = 1$). The regularized
model is found to have a smoother dynamical structure, but the main
features are similar for both cases indicating that our results are
not too sensitive to the particular method and amount of
regularization adopted.

In Figure~\ref{fig:dynmaj} we plot the dynamics in the `equatorial
cone' as function of radius. The upper panels show the rms velocities
for models A, B and C in the range between $r = 0.1''$ and $r=12''$
(corresponding to the regime where we have kinematic constraints along
the major axis). The middle panels display the ratio
$\sigma_a/\sigma_{\rm total}$ for the same radial interval
($\sigma_{\rm total}^2 = \sum_{a} \sigma_a^2$).  Since there is no
streaming motion in the radial and vertical directions, $\langle
\sigma_R \rangle$ and $\langle \sigma_z \rangle$ are equal to their
respective rms velocities. The difference between $\langle v_{\phi}^2
\rangle$ and $\sigma_{\phi}$ reflects the streaming motion of the
model. The lower panels show the ratios $\sigma_z/\sigma_R$ and
$\sigma_{\phi}/\sigma_R$. The three models differ predominantly in the
inner $\sim 3.0''$, reflecting the differences in BH mass, but are
very similar outside of $3.0''$: clearly $\langle v_{\phi}^2
\rangle$ dominates the dynamics outside this radius in accordance with
the rapid rotation of the outer disk. The rapid change in dynamical
structure (from azimuthally anisotropic to radially anisotropic) going
from $3''$ to $12''$ reflects the strong increase of $(V/\sigma)_{\rm
  obs}$ over this radial interval (see Figure~\ref{fig:velsig}): the
dynamically cold outer disk, mainly built up of close-to-circular
orbits with low $\sigma_{\phi}$, becomes the dominant mass component.
Over the same radial interval, the projected ellipticity increases
from $\sim 0.4$ to $\sim 0.7$, and this thus explains the observed
correlation between ellipticity and $V_{\rm obs}/V_{\rm mod}$ of the
best fitting isotropic Jeans model plotted in
Figure~\ref{fig:jeanscorr}.  At $R \gtrsim 8''$, $\sigma_{\phi}/\sigma_R
\sim 0.75$, not too different from the value in the solar neighborhood
where $\sigma_{\phi}/\sigma_R = 0.6$ (Dehnen 1998).

Models B and C have $\sigma_z / \sigma_R$ remarkably constant at $\sim
0.9$, and are thus not too different from two-integral models (for
which this ratio is exactly 1.0). This is very different from the
case without BH (model A) for which $\sigma_z / \sigma_R$ is a strong
function of radius $R$. This is the reason that the two-integral,
isotropic Jeans model discussed in Section~4, could not fit the HST
rotation velocities without BH, whereas model A can.
   
The main change going from model~A to model~C, is a strong increase of
$\sigma_{\phi}/\sigma_R$ in the inner $\sim 3''$. In these inner
regions the circular velocities increase strongly with increasing BH
mass.  Nevertheless, all three models provide an almost equally good
fit to the observed rotation velocities.  We checked our solutions and
found that going from model A to C, the ratio of $-L_z$-orbits over
$+L_z$-orbits increases in the radial interval $0.5'' < R < 3.0''$.
This causes the net streaming motions of all three models to be
roughly similar despite the large differences in circular velocities.
Furthermore, it explains the strong increase in $\sigma_{\phi}$
observed when going from model~A to model~C.

The dynamics of the bulge (represented by the cone with half-opening
angle of $60\deg$ centered around the minor axis) is presented in
Figure~\ref{fig:dynmin}. We only plot the results out to $R = 3.0''$,
corresponding to the radial interval where we have kinematic
constraints along the minor axis.  At $R \gtrsim 1''$ the models are
again similar, being dominated by rotation, albeit to a much
lesser extent than along the major axis.  Model~A is radially
anisotropic in the central region ($R \lesssim 1''$), and $\sigma_R$
decreases steadily with BH mass going from model~A to model~C. The
radial anisotropy of the central region of the bulge of model~A is
required to fit the high central velocity dispersions observed as good
as possible. Close to the equatorial plane, the steep rotation curve
observed prevents model~A from being too radially anisotropic (see
also Section~7.4).  Model~B is again remarkably close to two-integral
form, having $\sigma_z / \sigma_R \sim 1.0$.

\subsection{The influence of central radial anisotropy}

We now investigate to what extent radial anisotropy can influence the
central kinematics of NGC~4342. We solve the orbit weights for the
model with $\Upsilon_I = 7.25 \Msun/\Lsun$ and $M_{\rm BH} = 0$ using
two different sets of constraints. The first one uses all constraints
except the {\it rotation velocities} measured with the FOS. The second
one excludes only the FOS {\it velocity dispersions} from the
constraints. The results for both models are shown in
Figure~\ref{fig:radial}, where we plot the predictions for the
rotation velocities and velocity dispersions of both models. We only
plot the predictions for the FOS kinematics: both models have almost
indistinguishable WHT kinematics.  The lower two panels in
Figure~\ref{fig:radial} show the dynamics of the resulting models; we
plot both $\langle v_R^2\rangle^{1/2}$ and $\langle
v_{\phi}^2\rangle^{1/2}$ as function of radius. These quantities are
averaged between $\theta = 0\deg$ and $\theta = 90\deg$.

The model with the FOS rotation velocities excluded from the
constraints (solid lines) reveals the highest central velocity
dispersion achievable without a BH, while still fitting the WHT
measurements.  The model predicts a central rotation curve that is too
shallow to fit the observed rotation velocities, and is strongly
radially anisotropic in the center.  The model excluding the FOS
velocity dispersions (dashed lines) shows a very steep central
rotation curve.  The model is strongly azimuthally anisotropic in the
center, resulting in a much smaller central velocity dispersion.
Clearly, a model without a central BH cannot simultaneously fit the
rotation velocities and velocity dispersions measured with the FOS.

\subsection{The density distribution in the outer region of NGC~4342}

The best-fitting mass-to-light ratio of $\Upsilon_I = 6.3
M_{\odot}/L_{\odot}$ is unusually large for the stellar population of
an early-type galaxy (e.g., van der Marel 1991). The mass-to-light
ratio is mainly set by the large rotation velocities measured at the
outside of NGC~4342 (along the major axis). The large value found may
indicate that NGC~4342 is embedded in a massive dark halo. We did not
attempt to include such a component for two reasons: First, we are
mainly interested in trying to fit the central kinematics and to
examine the mass of a possible BH. The characteristics of a possible
dark halo will not affect the central dynamics, and therefore will not
alter our main conclusion that NGC~4342 harbors a massive BH. It may
however result in a different dynamical structure of the outer bulge
and disk.  Secondly, since we can fit all kinematics without having to
infer a dark halo, including such a component in the models will
merely lead to a large range of different models that can fit the
observed kinematics equally well.  Only kinematics at much larger
radii can constrain the presence and characteristics of such a
possible dark halo (cf.~R97).

Since mass-to-light ratio scales with distance $d$ as $d^{-1}$,
another explanation for the large value of $\Upsilon_I$ derived might
be an underestimate of the distance. In order to bring the inferred
mass-to-light ratio in accordance with the average value for
early-type galaxies, NGC~4342 has to be at about twice the assumed
distance of 15 Mpc. This is however hard to reconcile with the
observed heliocentric velocity of NGC~4342 of only $714 \kms$ (de
Vaucouleurs \etal 1976). Galaxies out to 15 Mpc from the center of the
Virgo cluster can still experience a Virgocentric infall of a few
hundred $\kms$: our local group falls towards Virgo with $\sim 200
\kms$ (Tammann \& Sandage 1985).  A distance of 30 Mpc for NGC~4342,
however, would imply a Virgocentric infall velocity of $\sim 1500
\kms$ (where we have taken Virgo to be at $1100 \kms$), which seems
too large.  So, the small recession velocity of NGC~4342 implies that
it can not be too far behind Virgo such as to have an appreciable
effect on the inferred mass-to-light ratio.

Another problem related to the outer density distribution is that the
mass model used (the MGE model) falls off as $\exp(-r^2)$ at large
radii. This is probably not correct, but since we only have photometry
that is limited to a small radial extent, we cannot examine this in
more detail.  The main requirement is that we have a proper mass model
for the major part of the galaxy. Our MGE model has a total luminosity
of $L_I = 3.57 \times 10^9 \Lsun$. We have shown in Section~3, that
for this luminosity we derive $B-V = 1.09$, in good agreement with the
average value for early-type galaxies. This therefore suggests that
our MGE model covers the major part of the galaxy, and that the galaxy
does not extend far beyond the radial extent of our MGE model.  In
addition, as is the case with a dark halo, a change in the luminosity
density profile in the outer parts of NGC~4342 will not affect our
main conclusions regarding the presence of a central BH.

\subsection{Comparison with other BH detections}

There are now about a dozen BH candidates known. The best cases are
the center of the MW, where a proper-motion study has revealed a $2.5
\times 10^6 \Msun$ BH (Eckart \& Genzel 1997), and NGC~4258, where
water masers are found to be in Keplerian rotation around a $3.6
\times 10^7 \Msun$ BH (Miyoshi \etal 1995).  Other strong evidence for
the existence of massive BHs comes from the observation of the Fe
K$\alpha$ line at 6.4 keV in a number of Seyfert 1 nuclei: the X-ray
emission line exhibits relativistic motions, interpreted as arising
from the accretion disk surrounding a massive BH (Tanaka \etal 1995;
Nandra \etal 1997).

Most other BH detections (or confirmations) are based on observations
with the HST. Observations of nuclear gas disks with the HST has
provided strong cases for massive BHs in M84, M87, NGC~4261, NGC~6251,
and NGC~7052. All other cases are based on stellar dynamical evidence:
M31, M32, NGC~3115, NGC~3377, NGC~3379, and NGC4486B. Of these only
M~32, NGC~3379 and NGC~4342 discussed here have been confronted with
three-integral modeling thus far.  As discussed in Section~1, one can
only hope to properly detect a BH if one can observe its kinematics
inside $r_{\rm BH}$.  All currently detected BHs have $r_{\rm BH} >
0.1''$, and since all these galaxies, except the nucleus of the Milky
Way, have now been observed with the HST, all current BH-candidates
fulfill this requirement.  Remarkable enough, most of these galaxies
(M31, M32, M87, NGC~3115, and NGC~4594) were already considered BH
candidates when the angular resolution of the observations was still
an order of magnitude larger, and similar to the angular size of the
radius of influence of the inferred BH (Rix 1993).  The BH in NGC~4342
has a relatively small radius of influence, with an angular size of
$\sim 0.4''$.  Indeed we have shown that only the HST/FOS observations
are of sufficient spatial resolution to rule out models without a
central BH.

Kormendy \& Richstone (1995) have suggested a relation between the
mass of the BH and that of the bulge of its parent galaxy: $\langle
M_{\rm BH}/M_{\rm bulge} \rangle = 0.22^{+0.14}_{-0.09} \times
10^{-2}$. For more recent discussions on this relation, including more
recent BH detections see Kormendy \etal (1998), Ho (1998), van der
Marel (1998), and van der Marel \& van den Bosch (1998).  NGC~4342 is
the galaxy with currently the second-highest ratio of BH mass over
bulge mass of $2.6^{+1.5}_{-0.9} \times 10^{-2}$, and is only
superseded by the peculiar galaxy NGC~4486B, for which the case of a
massive BH is less strong (Kormendy \etal 1997).  NGC~3115 ($M_{\rm
  BH}/M_{\rm bulge} = 2.4 \times 10^{-2}$) and the Milky Way ($M_{\rm
  BH}/M_{\rm bulge} = 0.017 \times 10^{-2}$) are other clear deviants.
Clearly, the scatter around the suggested relation is considerable,
seemingly as large as nearly two orders of magnitude.

\section{Conclusions}

Spectra obtained with the WHT and HST/FOS of the edge-on S0 galaxy
NGC~4342 have revealed a very steep central rotation curve and a
strong central increase in velocity dispersion. These data suggest a
large central mass concentration. In this paper we presented detailed
dynamical models of NGC~4342 used to investigate whether its nucleus
harbors a massive BH.

We model the luminous density distribution of NGC~4342 with multiple
Gaussian components. After projection and PSF convolution this model
provides an excellent fit to the HST $I$-band surface brightness
distribution. The parameters of this model were derived with the MGE
method.

Simple isotropic Jeans models suggest that NGC~4342 harbors a massive
BH of a few times $10^8 \Msun$. The actual mass of the BH depends on
the data-set fitted: the WHT data suggest $M_{\rm BH} \approx 3 \times
10^8 \Msun$; the HST/FOS data suggest a somewhat larger BH mass of
$\sim 6 \times 10^8 \Msun$. This discrepancy already suggests that the
assumptions underlying the Jeans models, i.e., $f=f(E,L_z)$ and
therefore $\sigma_R = \sigma_z$, are incorrect. This is also evident
from the fact that the Jeans models cannot accurately fit the
major-axis rms velocities measured with the WHT. These rms velocities
are independent of the freedom in the anisotropy
$\sigma_{\phi}/\sigma_{R}$ allowed in the Jeans modeling. We find that
for a mass-to-light ratio $\Upsilon_I = 6.2 \Msun/\Lsun$ and an
isotropic velocity distribution ($\sigma_R = \sigma_z =
\sigma_{\phi}$) the Jeans model with $M_{\rm BH} = 3 \times 10^8
\Msun$ provides the best fit to the observed WHT velocity dispersions
along the major axis.  However, the rotation velocities are not very
well fitted and we find a correlation between $V_{\rm obs}/V_{\rm
  mod}$ and the local ellipticity of the projected surface brightness,
such that the model underpredicts the rotation velocities in the
highly flattened regions (dominated by the disk light) and
overpredicts them in the less flattened region (dominated by the bulge
light). This suggests that the different components in NGC~4342 have
different velocity anisotropies.

We thus constructed three-integral axisymmetric models of NGC~4342 in
order to examine the mass of a possible BH and the dynamical structure
of the different components. The modeling technique is an extension of
Schwarzschild's orbit-superposition technique, and is based on finding
the ensemble of orbits that best fits the observations. These models
make no assumption about the dynamical structure and are fully
general. This technique, developed by Rix \etal (1997) and Cretton
\etal (1998), has previously been used to prove the existence of a
massive BH of $(3.4 \pm 0.7) \times 10^6 \Msun$ in the compact
elliptical M32 (van der Marel \etal 1998). We have constructed a range
of dynamical ($M_{\rm BH},\Upsilon_I$)-models of NGC~4342 to determine
a central BH mass of $3.0^{+1.7}_{-1.0} \times 10^8 \Msun$ and an
$I$-band mass-to-light ratio of $6.3^{+0.5}_{-0.4} \Msun/\Lsun$. The
high spatial resolution of the HST/FOS data allow us to rule out
models without a BH to a confidence level better than $99.73$ per
cent.  With a similar confidence we can rule out models with a BH more
massive than $7 \times 10^8 \Msun$. This upper limit on the BH mass is
mainly due to the VP shape parameters $h_3$ and $h_4$.  With the
current data we can not rule out alternatives to a massive BH, such as
a cluster of brown dwarfs or stellar remnants.  Nevertheless, the QSO
paradigm, together with the fact that the presence of massive BHs in
galactic nuclei has unambiguously been demonstrated in a few galaxies
were the inferred central densities are high enough to rule out dark
clusters as alternatives (see discussion in Maoz 1997), make the
interpretation of the inferred MDO in NGC~4342 in terms of a massive
BH the most likely.

We computed the intrinsic mean velocities and velocity dispersions of
the three-integral models. The dynamical structures of the best
fitting models vary strongly with radius, reflecting the
multi-component structure of NGC~4342. Between $2''$ and $12''$ in the
equatorial plane the best fitting models change from azimuthally
anisotropic to radially anisotropic, while $\sigma_z / \sigma_R
\approx 0.9$. This explains the correlation between the projected
ellipticity and the failure of the isotropic Jeans models to fit the
observed rotation velocities along the major axis. The bulge in the
best fitting model without BH is radially anisotropic. However, we
have shown that even without the constraints of the measured HST/FOS
rotation velocities, models without BH cannot fit the central HST/FOS
velocity dispersion. The rotation velocities measured from the ground
already constrain the amount of central radial anisotropy such that
models without a BH cannot fit the high central velocity dispersion
measured with the HST/FOS.

The BH mass thus derived contributes a fraction of $2.6^{+1.5}_{-0.9}$
per cent to the total mass of the bulge ($1.2 \times 10^{10} \Msun$).
With this BH mass, NGC~4342 has one of the highest ratios of BH mass
over bulge mass. Currently, the BH in our own galaxy has, with $0.02$
per cent, the lowest BH mass to bulge mass ratio known: the scatter in
the $M_{\rm BH}$ vs. $M_{\rm bulge}$ relation seems to be as large as
two orders of magnitude.  Extremely high spatial resolution is
required in order to investigate if other galaxies have even lower
values of $M_{\rm BH}/M_{\rm bulge}$. In conclusion, current data are 
consistent with a relation between bulge mass and BH mass, but
the scatter is very large, and it is likely that the current $M_{\rm
  BH}/M_{\rm bulge}$ ratios found are biased towards an upper limit.
Although the newly installed Space Telescope Imaging Spectrograph
(STIS) is likely to detect many more BH cases in the coming years,
detection of BHs with masses of the order of a $0.02$ per cent of the
bulge mass or less in galaxies in Virgo or beyond, will probably have
to await a next generation space telescope.

%%%%%%%%%%%%%%%
% Acknowledgments
%%%%%%%%%%%%%%%

\acknowledgments

Part of the calculations presented in this paper were performed on
two UltraSparcs, kindly made availably by the Lorentz Center. Many
thanks are due to Eric Emsellem, for his help with the MGE analysis,
and to Tim de Zeeuw, Walter Jaffe, HongSheng Zhao and Roeland van der
Marel for many fruitful discussions. FvdB was supported by a Hubble
Fellowship, \#HF-01102.11-97A, awarded by STScI. NC acknowledges the
hospitality of Steward Observatory where part of this work was done. 

\clearpage
 
%%%%%%%%%%%%%%%
% Use a small baselineskip for the references, unless in submission mode.
%%%%%%%%%%%%%%%

\ifsubmode\else
\baselineskip=10pt
\fi

%%%%%%%%%%%%%%%
% Reference List
%%%%%%%%%%%%%%%

\clearpage

%%%%%%%%%%%%%%%
% Change back to the regular baselineskip, if necessary
%%%%%%%%%%%%%%%

\ifsubmode\else
\baselineskip=14pt
\fi

%%%%%%%%%%%%%%%
% Figure Captions
%%%%%%%%%%%%%%%

\figcaption[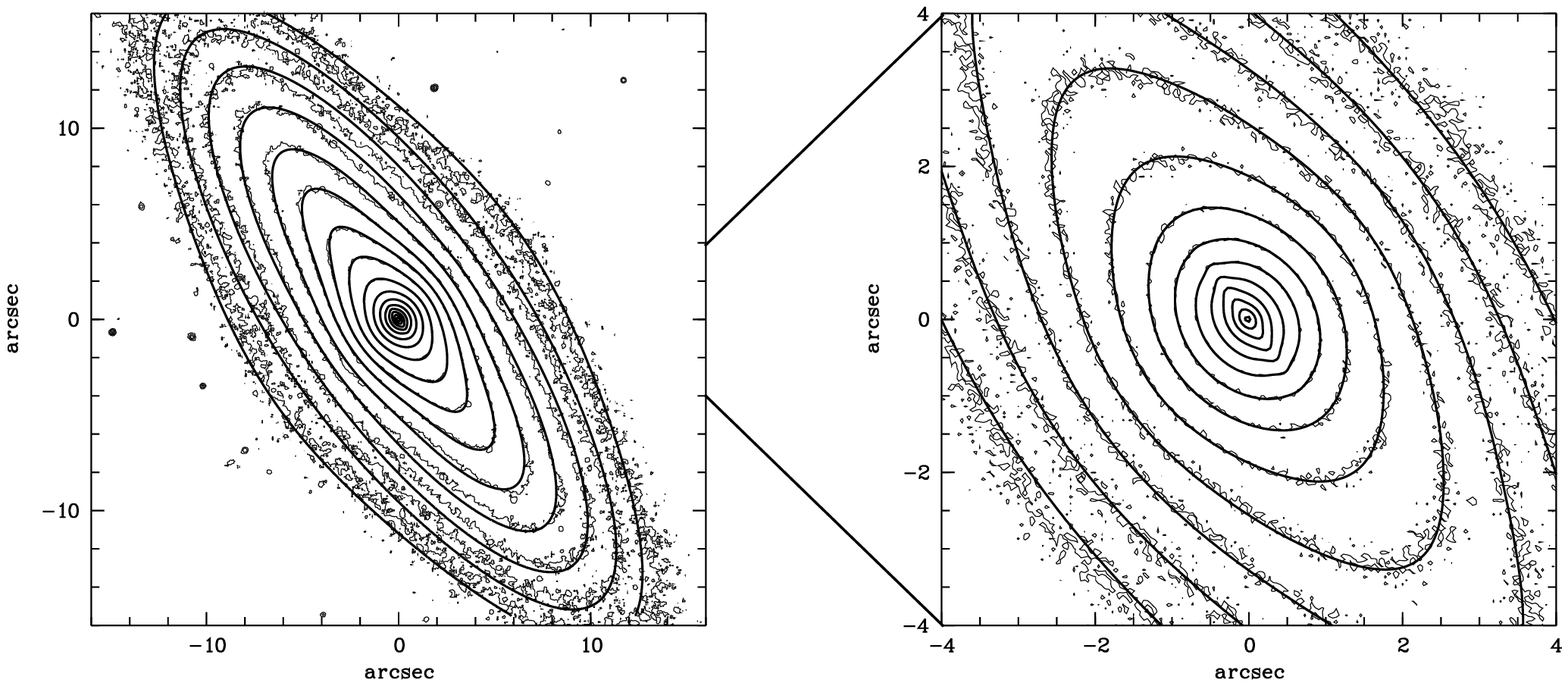]{Contour maps of the WFPC2 $I$-band image of
  NGC~4342 at two different scales: $32''\times 32''$ (left-hand
  panel) and $8'' \times 8''$ (right-hand panel). Superimposed are the
  contours of the MGE model of the intrinsic surface brightness
  convolved with the HST PSF (see Section~3).\label{fig:mge}}

\figcaption[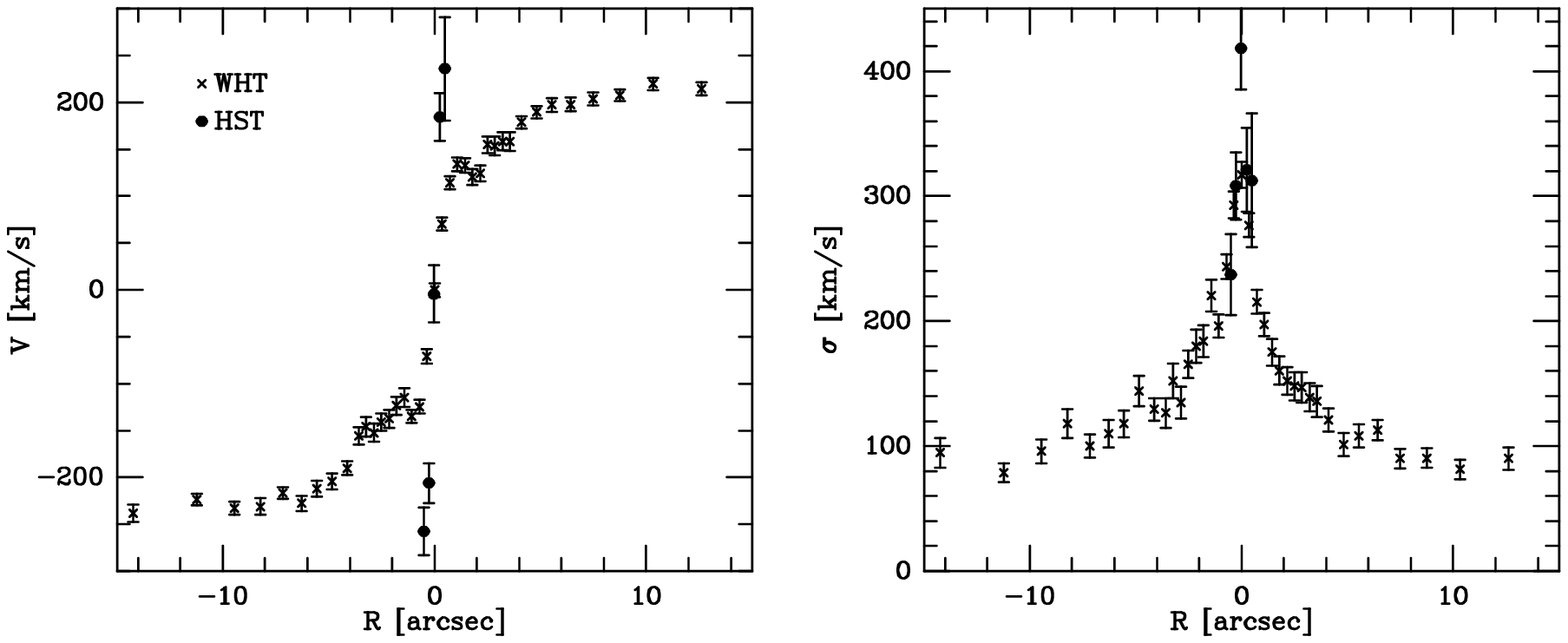]{Observed rotation velocities $V$ and
  velocity dispersions $\sigma$ (as determined from the best-fitting
  Gaussian, see text) along the major axis of NGC~4342 obtained with
  the WHT (crosses) and the FOS (solid dots).  The gradient of the
  rotation velocity and the central velocity dispersion increase
  considerably going to the four times higher spatial resolution of
  the FOS. See BJM98 for details on the data.\label{fig:kinemdat}}

\figcaption[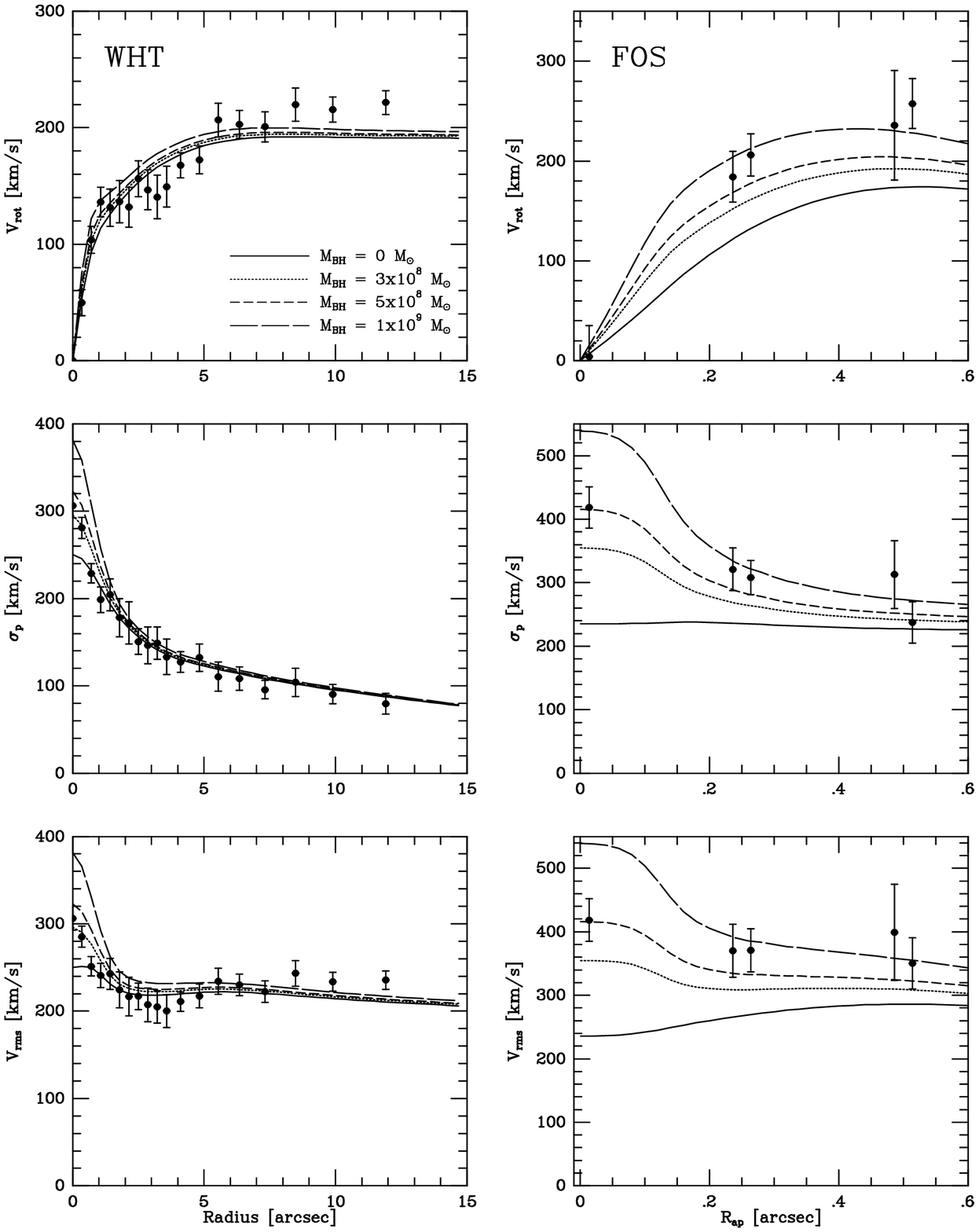]{Results of the Jeans modeling. The solid
  dots with errorbars indicate the observed rotation velocities and
  velocity dispersions.  Overplotted are four models that differ only
  in the mass of the central BH (0, 3, 5 and $10 \times 10^8 \Msun$).
  All models have fully isotropic velocity dispersions, and a stellar
  mass-to-light ratio of $\Upsilon_I = 6.2 \Msun/\Lsun$. The three
  panels on the left show the {\tt WHT} kinematics and the model
  predictions for $V_{\rm rot}$, $\sigma_p$ and $V_{\rm rms}$. The
  model with $M_{\rm BH} = 3 \times 10^8 \Msun$ provides the best fit
  to the velocity dispersions. Neither of the four models provides an
  accurate fit to the rotation velocities (see also 
  Figure~\ref{fig:jeanscorr}). The
  panels on the right compare the model predictions with the HST/FOS
  kinematics. Both the rotation velocities and the velocity
  dispersions suggest the presence of a BH with a mass of $\sim 6
  \times 10^8 \Msun$.\label{fig:jeansmod}}

\figcaption[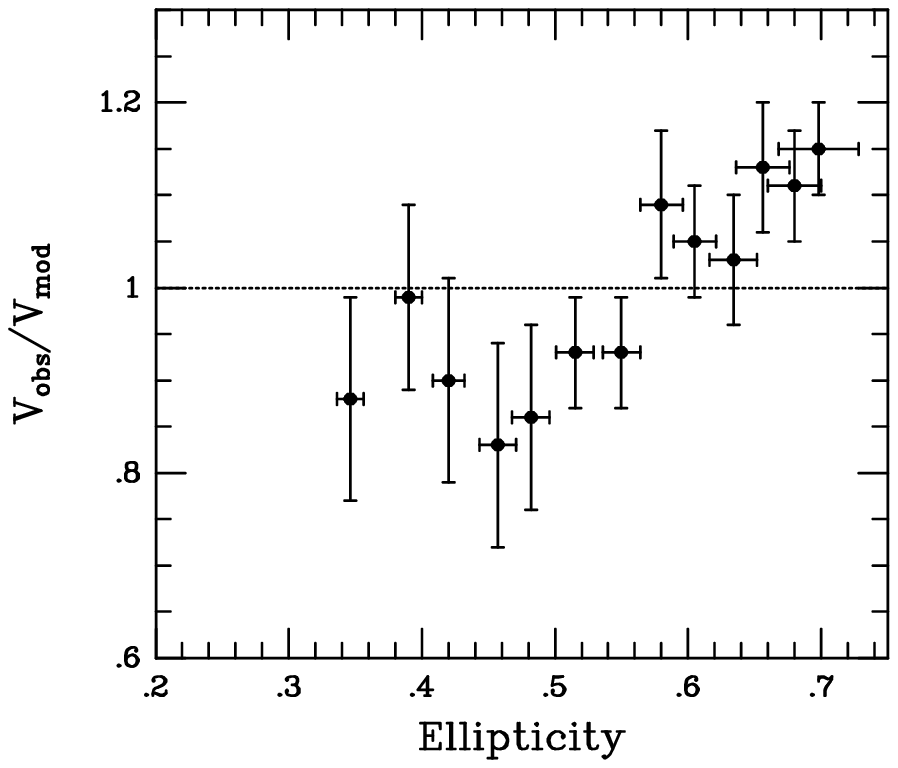]{The ratio $V_{\rm obs}/V_{\rm mod}$ of the
  observed rotation velocities over the rotation velocities predicted
  by the isotropic Jeans model with $\Upsilon_I = 6.2 \Msun/\Lsun$ and
  $M_{\rm BH} = 3 \times 10^8 \Msun$, as function of the local,
  projected ellipticity of the isophotes. Only results beyond $2''$
  are shown. At smaller radii seeing influences the observed
  quantities significantly. There is a clear correlation in that the
  Jeans model underpredicts the rotation velocities in the moderately
  flattened region, and overpredicts the rotation velocities in the
  radial interval where the isophotes are strongly flattened, and thus
  dominated by the light of the outer disk
  component.\label{fig:jeanscorr}}

\figcaption[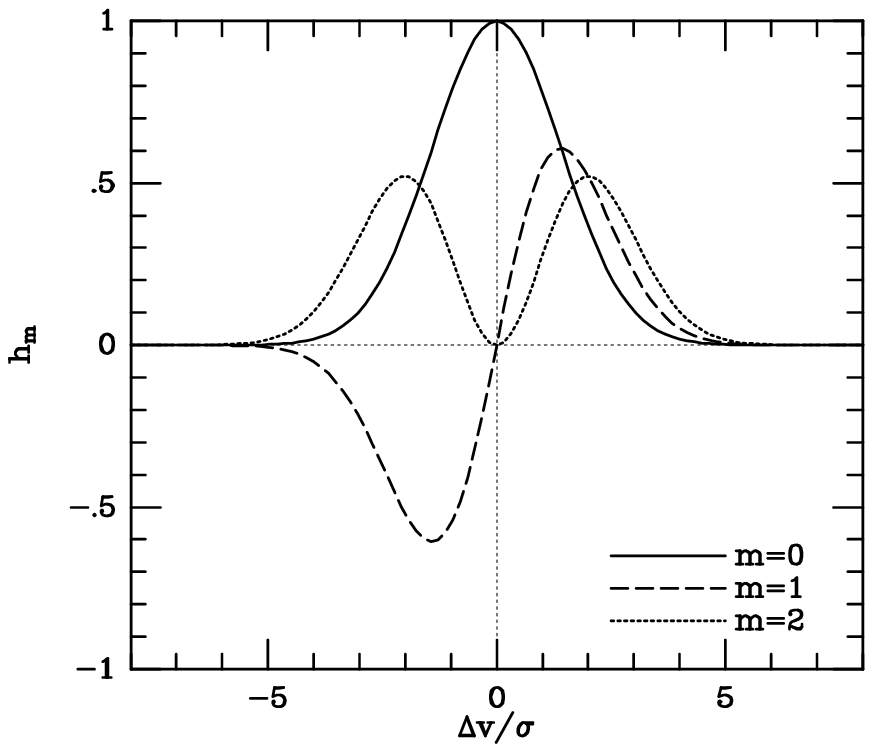]{The Gauss-Hermite moments $h_m$ $(m=0,1,2)$ of a
  Gaussian VP expanded around another Gaussian with the same
  dispersion $\sigma$. Results are plotted as functions of the
  velocity difference $\Delta V$ between the two Gaussians, expressed
  in units of $\sigma$.  For $\Delta V = 0$, the two Gaussians are
  identical and $h_0 =1$ and $h_m = 0$ ($m=1,2,3,...$).\label{fig:gh}}

\figcaption[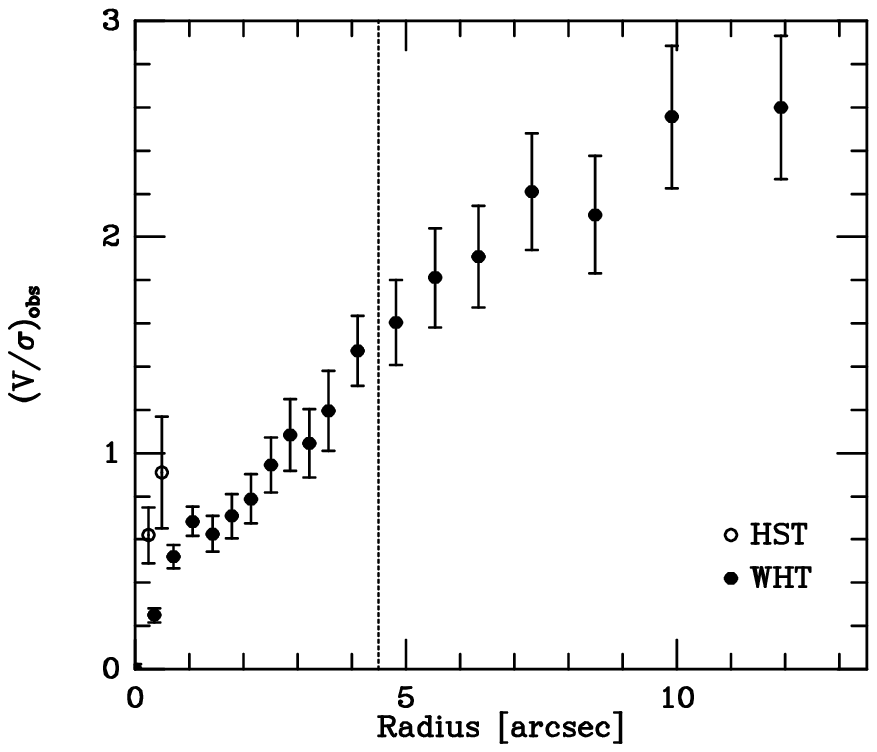]{The ratio $V/\sigma$ as function of radius of
  the major-axis kinematics of NGC~4342 as measured with the WHT
  (solid dots) and the HST/FOS (open circles). The dotted line
  indicates $R_{\rm lim}$ outside which $V/\sigma > 1.5$. In the
  modified approach, we exclude counter-rotating orbits with circular
  radii $R_c > R_{\rm lim}$ from the NNLS.\label{fig:velsig}}

\figcaption[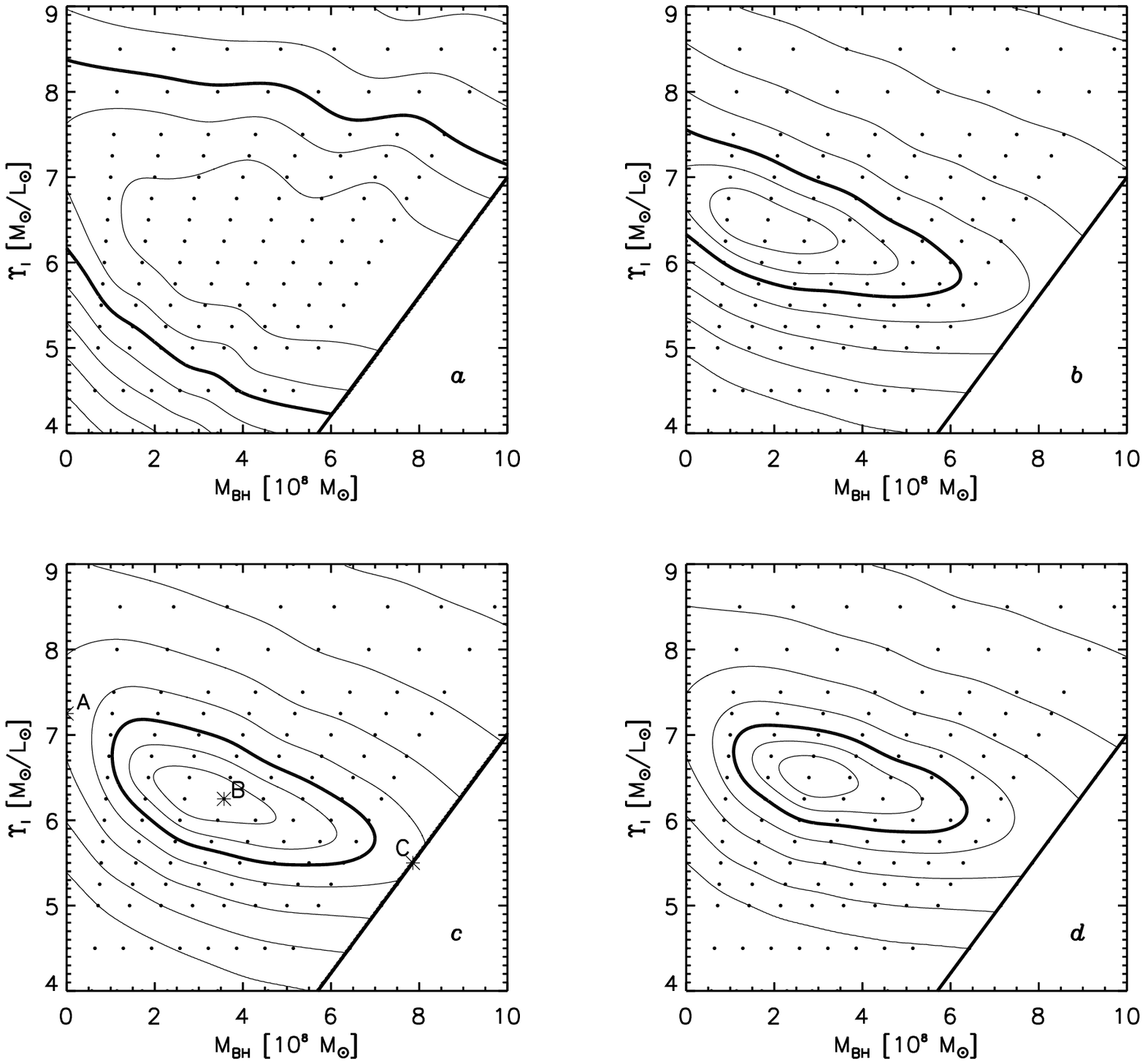]{Contour plots of $\chi^2 = \chi^2_{\rm
    obs} + \chi^2_{\rm sc}$, which measures the goodness-of-fit to the
  constraints as function of BH mass, $M_{\rm BH}$, and $I$-band
  mass-to-light ratio, $\Upsilon_I$. The first three contours define
  the formal $68.3$, $95.4$ and $99.73$ per cent confidence levels
  (latter one is plotted with a thick contour). The subsequent
  contours are characterized by a factor two increase in $\Delta
  \chi^2$. Solid dots indicate actual model calculations.  The
  $\chi^2$ surface of panel $a$ results from excluding $h_3$ and $h_4$
  as well as all HST/FOS measurements from the constraints.  In panel
  $b$ we have only excluded the HST/FOS data, and panel $c$ shows the
  resulting $\chi^2$ plot when all constraints are taken into account.
  The asterisks labeled A, B and C indicate special models discussed
  in the text. Finally, in panel $d$ we have used the modified
  approach and included the additional constraint that
  counter-rotating orbits with $R_c > 4.5''$ are not allowed in the
  solution.\label{fig:chicont}}

\figcaption[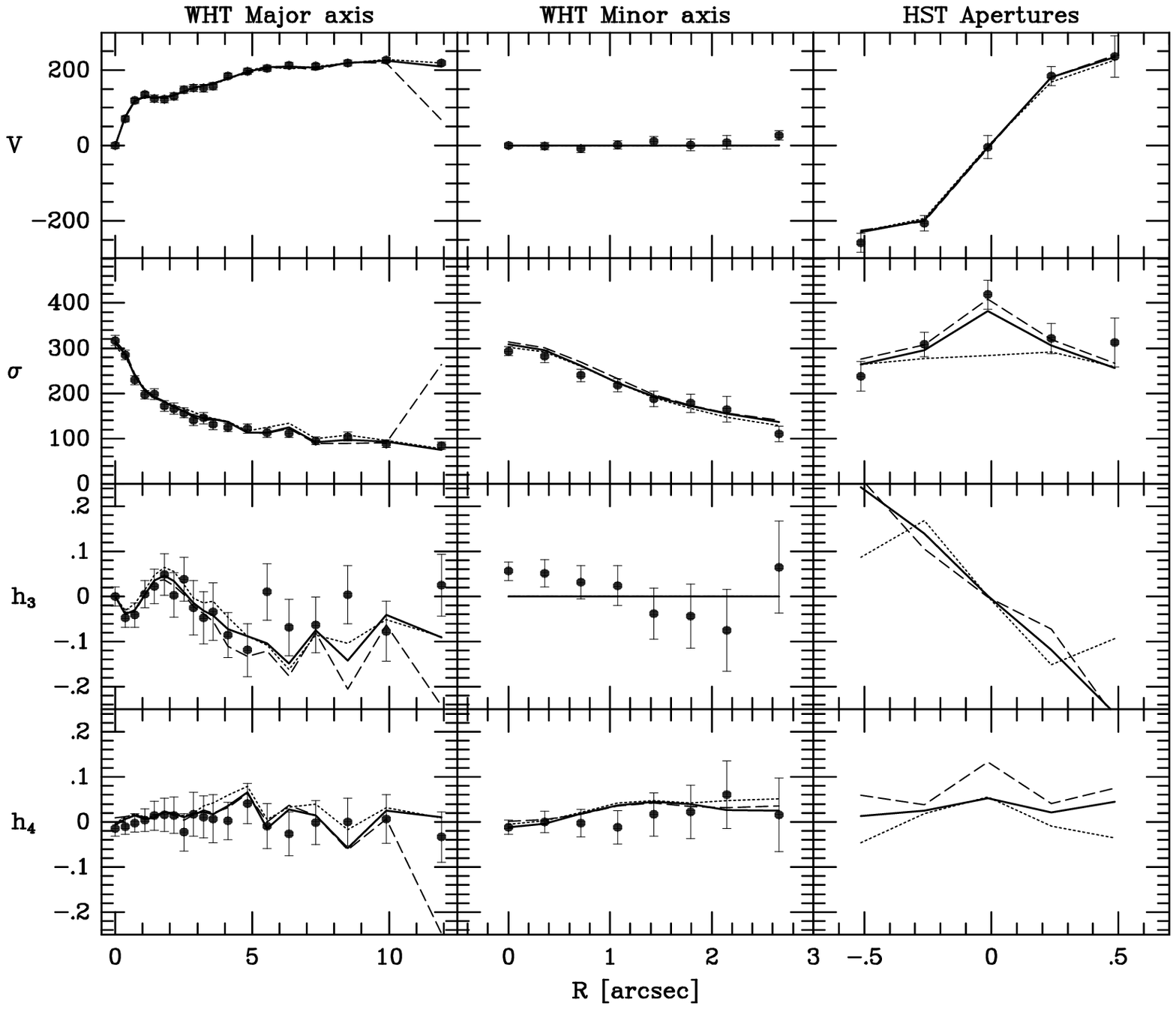]{Kinematics for the three data sets
  (solid dots with errorbars) compared to the predictions for models A
  (dotted lines), B (solid lines) and C (dashed lines). Although we
  did not measure $h_3$ and $h_4$ from our HST/FOS spectra, we plot
  the model predictions in the lower two panels on the right.  Model
  A, which has no BH, strongly underestimates the central velocity
  dispersion as measured with the FOS. The strong deviation of the
  predictions of model C for the outermost WHT-point along the major
  axis, is due to the `artificial' counter-rotation of the model (see
  Figure~\ref{fig:doublepeak}).\label{fig:modelfit}}

\figcaption[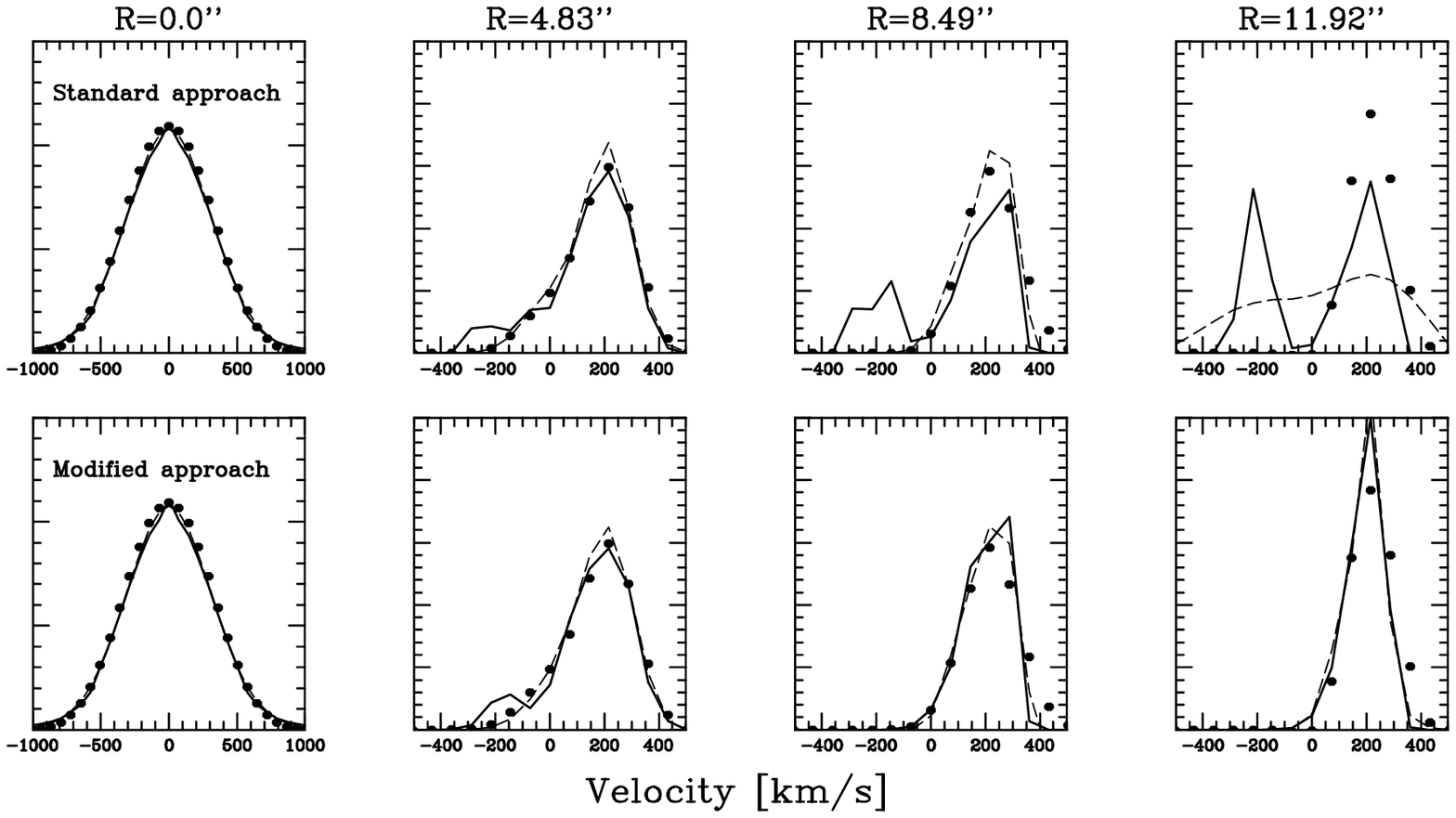]{Velocity profiles at four different radii along
  the major axis of NGC~4342. The observed VPs, reconstructed from the
  measured $V$, $\sigma$, $h_3$ and $h_4$ while assuming $h_m = 0$ for
  $m \geq 5$, are plotted as solid dots. The solid lines correspond to
  the model VPs, and the dashed lines to the Gauss-Hermite series (up
  to order 4) fitted to these model VPs. Upper panels correspond to
  model C, whereas lower panels correspond to a model with the same BH
  mass and mass-to-light ratio, but whose orbit solution was computed
  with the modified approach.  The upper panels clearly reveal that
  the model VPs at larger radii are double peaked. This can yield
  strange Gaussian fits, as evident from the upper right panel. The
  modified approach excludes counter-rotating orbits with $R_c >
  4.5''$ from the orbit library, and does not reveal these double
  peaked VPs.\label{fig:doublepeak}}

\figcaption[chi_eps.ps]{The value of $\chi^2 = \chi^2_{\rm obs} +
  \chi^2_{\rm sc}$, which measures the goodness-of-fit to the
  constraints, as function of the scale length $\epsilon$ (in arcsec)
  of the Plummer potential with a total mass of $3.6 \times 10^8
  \Msun$ representing a cluster of dark objects in the center of
  NGC~4342.  All models have a stellar mass-to-light ratio of $6.25$.
  The best fit to the data is achieved in the limit $\epsilon
  \rightarrow 0$, which corresponds to our model with a BH rather than
  a dark cluster.  We can rule out models with $\epsilon > 0.07''$ at
  the 99.73 per cent confidence level.\label{fig:chieps}}

\figcaption[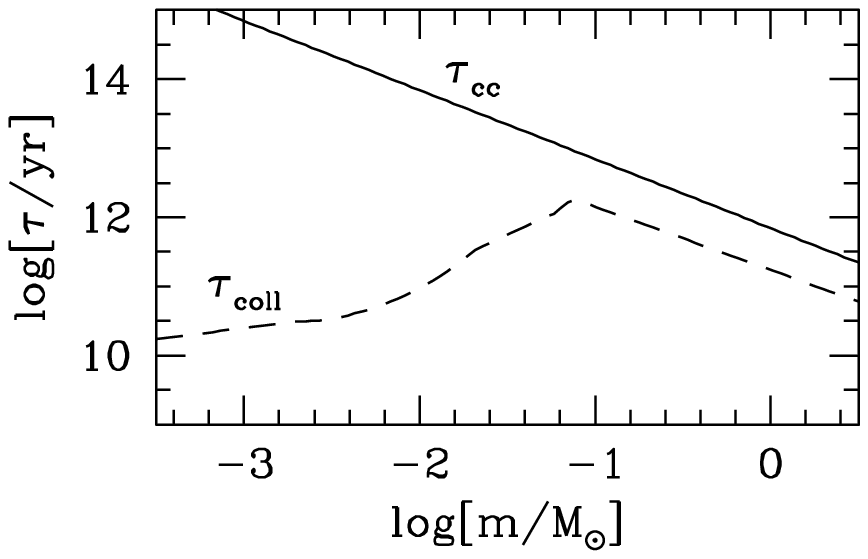]{The timescales for core collapse ($\tau_{\rm
    cc}$) and collisional destruction ($\tau_{\rm coll}$) of a $3.6
  \times 10^8 \Msun$ Plummer cluster of non-collapsed objects with a
  central density of $6.7 \times 10^5 \Msun {\rm pc}^{-3}$ as function
  of the mass $m$ of the constituents. If any of these timescales is
  less than $\sim 10^{10}$ years, it makes such cluster an unlikely
  alternative to a BH. Unfortunately, current data does not allow us
  to rule out any of the dark clusters as an alternative MDO in
  NGC~4342.\label{fig:lifetime}}

\figcaption[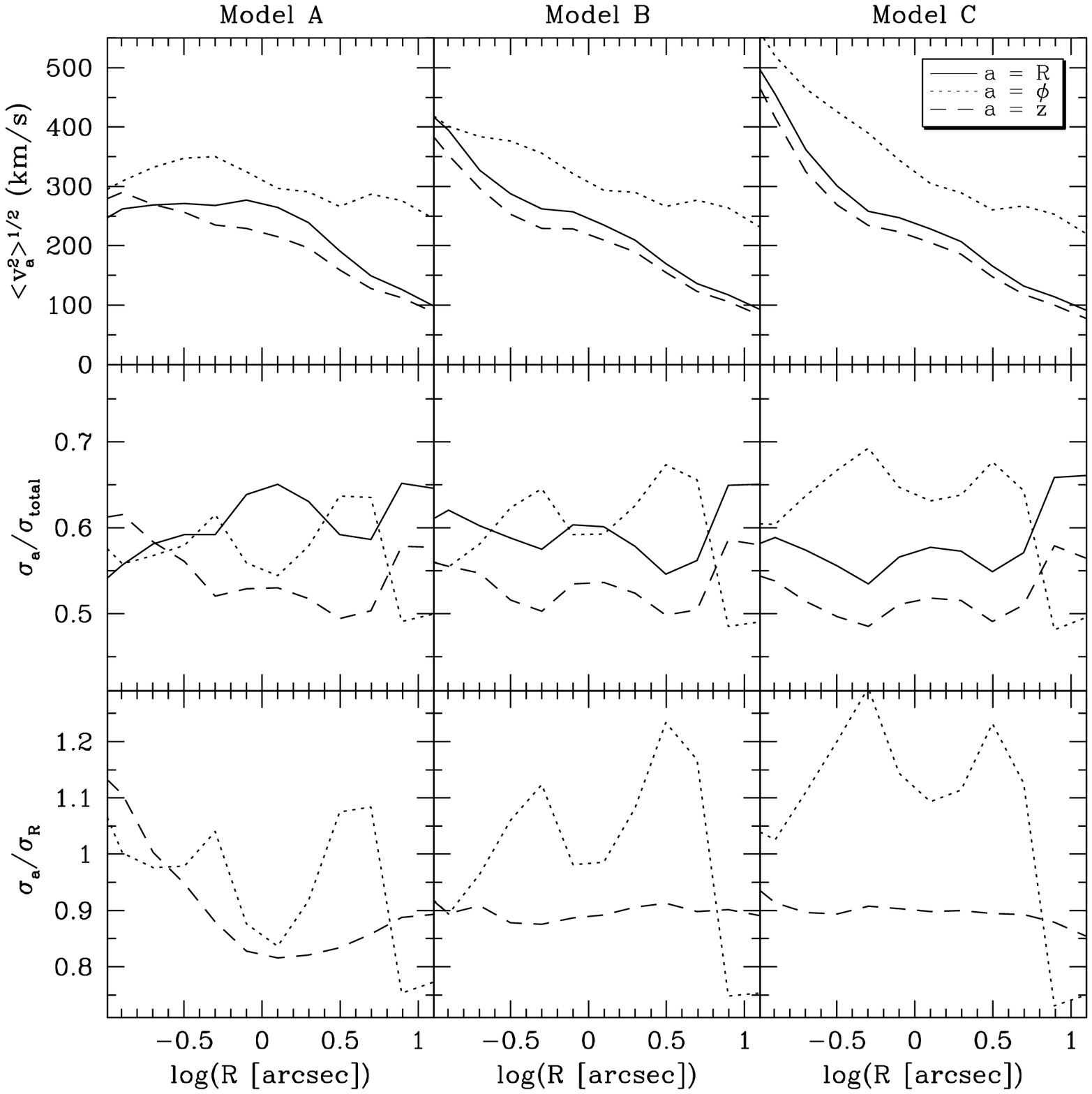]{The dynamical structure of models A, B
  and C averaged over a cone with half-opening angle of $30\deg$
  centered on the equatorial plane. Upper panels show the rms
  velocities $\langle v_a^2 \rangle^{1/2}$ in $\kms$ , middle panel
  the normalized velocity dispersions $\sigma_a/\sigma_{\rm total}$,
  and lower panels the ratios $\sigma_a/\sigma_R$.  Solid curves are
  for the radial component ($a=R$), dotted curves for the azimuthal
  component ($a=\phi$), and dashed curves for the vertical component
  ($a=z$).  Results are plotted over the radial interval where we have
  kinematic constraints along the major axis, i.e., $0.1'' < R <
  12''$.\label{fig:dynmaj}}

\figcaption[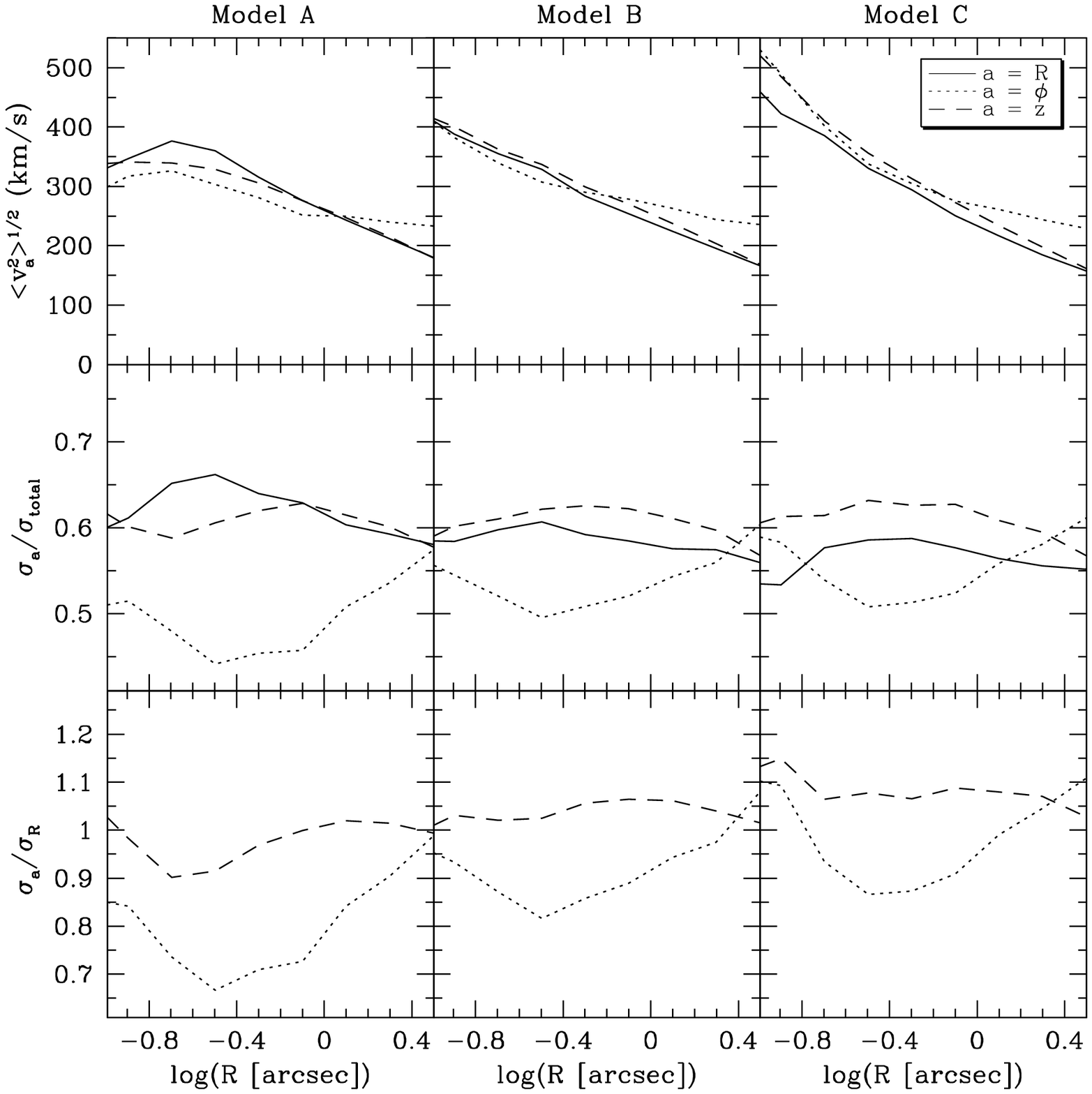]{Same as Figure~\ref{fig:dynmaj}, except
  that here the quantities are averaged over a cone with half-opening
  angle of $60 \deg$ centered on the symmetry axis $R=0$) of the
  models.\label{fig:dynmin}}

\figcaption[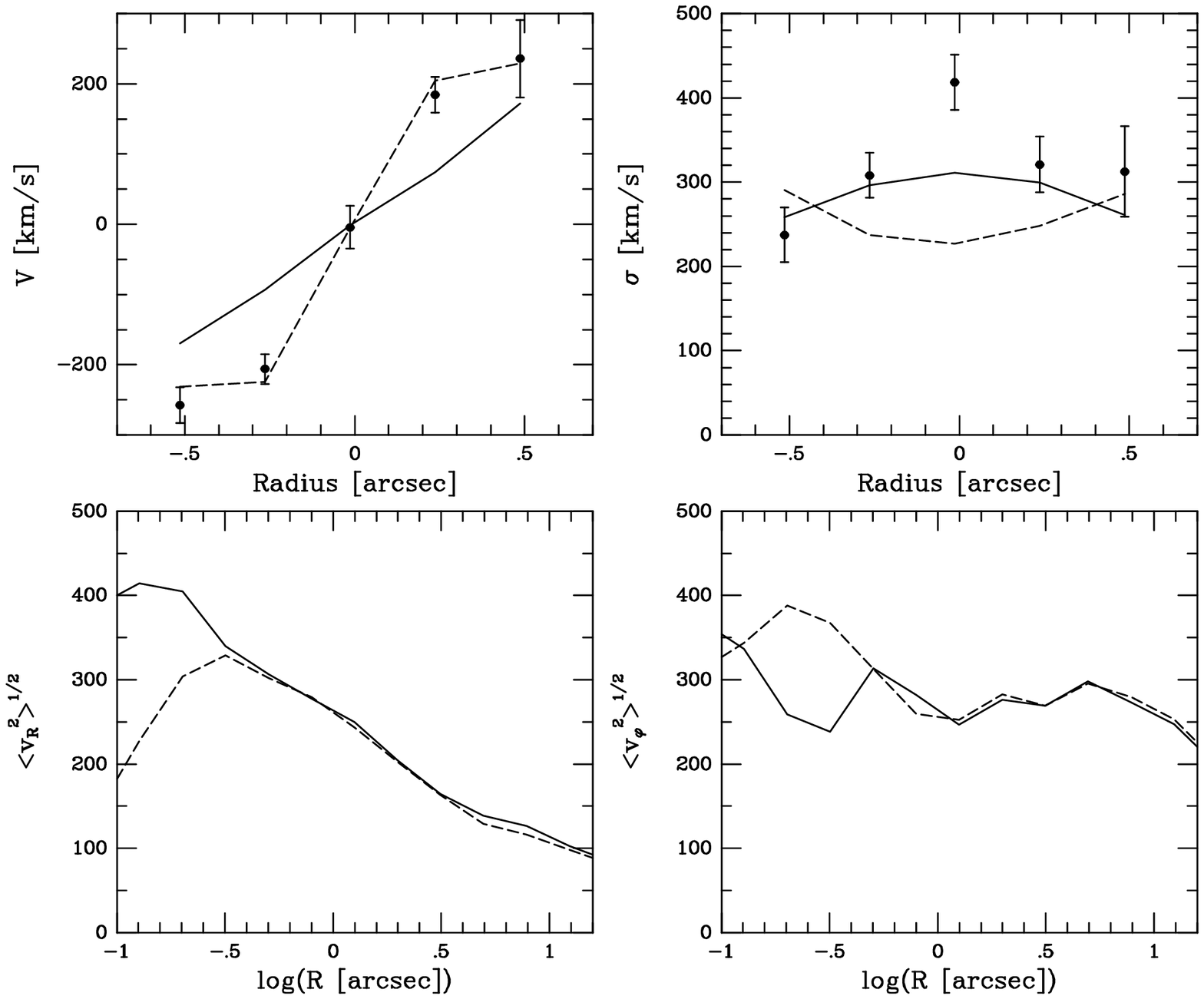]{Radial velocity anisotropy at work: Solid lines
  are for a model with $\Upsilon_I = 7.25 \Msun/\Lsun$ and $M_{\rm BH}
  = 0$, where we have neglected the FOS measurements of the rotation
  velocities. The dashed lines correspond to the same model, but for
  which we have neglected the FOS velocity-dispersion measurements.
  The two lower panels show the intrinsic dynamics, $\langle
  v_R^2\rangle^{1/2}$ (lower-left panel) and $\langle
  v_{\phi}^2\rangle^{1/2}$ (lower-right panel), of the two
  models.\label{fig:radial}}

%%%%%%%%%%%%%%%
% Print Figures (Only if printfigtrue)
%%%%%%%%%%%%%%%

\ifprintfig

\clearpage
\begin{figure}
\epsfxsize=15.0truecm
\centerline{\epsfbox{totfig.eps}}
\vskip3.0truecm
\addtocounter{figure}{-13}
\centerline{Figure~\thefigure}
\end{figure}

\clearpage
\begin{figure}
\epsfxsize=15.0truecm
\centerline{\epsfbox{velsig4342.ps}}
\vskip3.0truecm
\addtocounter{figure}{1}
\centerline{Figure~\thefigure}
\end{figure}

\clearpage
\begin{figure}
\epsfxsize=14.0truecm
\centerline{\epsfbox{res_jeans3.ps}}
\vskip3.0truecm
\addtocounter{figure}{1}
\centerline{Figure~\thefigure}
\end{figure}

\clearpage
\begin{figure}
\epsfxsize=12.0truecm
\centerline{\epsfbox{satoh.ps}}
\vskip3.0truecm
\addtocounter{figure}{1}
\centerline{Figure~\thefigure}
\end{figure}

\clearpage
\begin{figure}
\epsfxsize=12.0truecm
\centerline{\epsfbox{gh2.ps}}
\vskip3.0truecm
\addtocounter{figure}{1}
\centerline{Figure~\thefigure}
\end{figure}

\clearpage
\begin{figure}
\epsfxsize=12.0truecm
\centerline{\epsfbox{velsig.ps}}
\vskip3.0truecm
\addtocounter{figure}{1}
\centerline{Figure~\thefigure}
\end{figure}

\clearpage
\begin{figure}
\epsfxsize=15.0truecm
\centerline{\epsfbox{allchi2.ps}}
\vskip3.0truecm
\addtocounter{figure}{1}
\centerline{Figure~\thefigure}
\end{figure}

\clearpage
\begin{figure}
\epsfxsize=15.0truecm
\centerline{\epsfbox{kinemresults3.ps}}
\vskip3.0truecm
\addtocounter{figure}{1}
\centerline{Figure~\thefigure}
\end{figure}

\clearpage
\begin{figure}
\epsfxsize=15.0truecm
\centerline{\epsfbox{vps.ps}}
\vskip3.0truecm
\addtocounter{figure}{1}
\centerline{Figure~\thefigure}
\end{figure}

\clearpage
\begin{figure}
\epsfxsize=15.0truecm
\centerline{\epsfbox{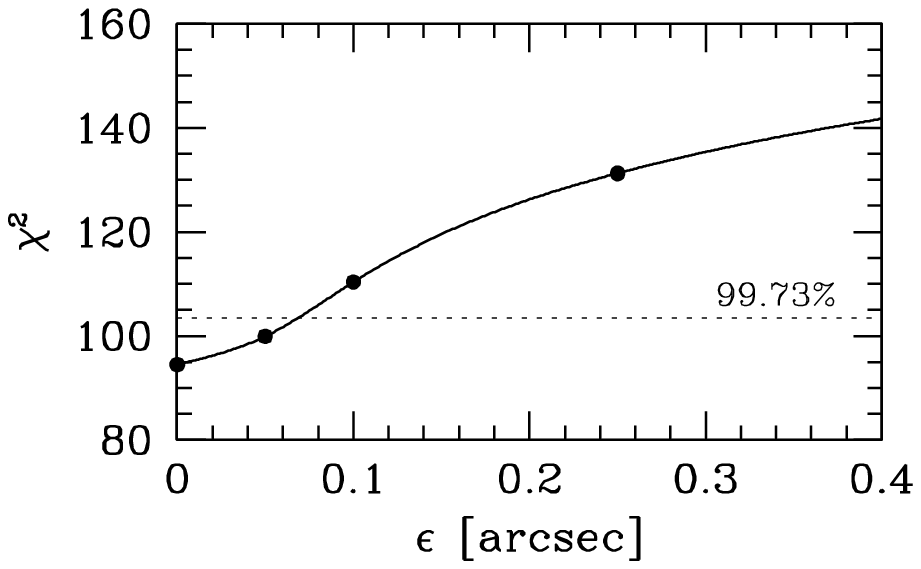}}
\vskip3.0truecm
\addtocounter{figure}{1}
\centerline{Figure~\thefigure}
\end{figure}

\clearpage
\begin{figure}
\epsfxsize=15.0truecm
\centerline{\epsfbox{lifetime.ps}}
\vskip3.0truecm
\addtocounter{figure}{1}
\centerline{Figure~\thefigure}
\end{figure}

\clearpage
\begin{figure}
\epsfxsize=15.0truecm
\centerline{\epsfbox{kinem_maj.ps}}
\vskip3.0truecm
\addtocounter{figure}{1}
\centerline{Figure~\thefigure}
\end{figure}

\clearpage
\begin{figure}
\epsfxsize=15.0truecm
\centerline{\epsfbox{kinem_min.ps}}
\vskip3.0truecm
\addtocounter{figure}{1}
\centerline{Figure~\thefigure}
\end{figure}

\clearpage
\begin{figure}
\epsfxsize=15.0truecm
\centerline{\epsfbox{radial.ps}}
\vskip3.0truecm
\addtocounter{figure}{1}
\centerline{Figure~\thefigure}
\end{figure}

\fi

%%% END OF FIGURES %%%

%%%%%%%%%%%%%%%
% Tables 
%%%%%%%%%%%%%%%

\clearpage
\ifsubmode\pagestyle{empty}\fi

%%% TABLE 1 %%%

\begin{deluxetable}{crrrr}
\tablecaption{Parameters of MGE model for the deconvolved
$I$-band surface brightness. \label{tbl-1}}
\tablehead{
\colhead{$j$} & 
\colhead{$I'_j$} & 
\colhead{$a'_j$} & 
\colhead{$q'_j$} & 
\colhead{$L_{I,j}$} \\
\colhead{(1)} &
\colhead{(2)} &
\colhead{(3)} &
\colhead{(4)} &
\colhead{(5)} \\
} 
\startdata
 1 & 490833.0 &  0.032 & 0.817 & $1.40 \times 10^7$ \\
 2 &  99417.9 &  0.101 & 0.865 & $2.92 \times 10^7$ \\ 
 3 &  67415.3 &  0.282 & 0.601 & $1.07 \times 10^8$ \\
 4 &  84108.1 &  0.343 & 0.136 & $4.47 \times 10^7$ \\ 
 5 &  28511.8 &  0.394 & 0.856 & $1.26 \times 10^8$ \\ 
 6 &  15529.3 &  0.753 & 0.622 & $1.82 \times 10^8$ \\ 
 7 &   8490.8 &  0.756 & 1.000 & $1.61 \times 10^8$ \\ 
 8 &   6055.4 &  1.866 & 0.665 & $4.66 \times 10^8$ \\ 
 9 &   2951.4 &  4.419 & 0.250 & $4.79 \times 10^8$ \\ 
10 &   1572.8 &  9.229 & 0.266 & $1.18 \times 10^9$ \\ 
11 &    229.9 & 11.854 & 0.723 & $7.76 \times 10^8$ \\ 
\enddata
\tablecomments{Column~(1) gives the index number of each Gaussian. 
  Column~(2) gives its central surface brightness; column~(3) its
  standard deviation (which expresses the size of the Gaussian along
  the major axis); column~(4) its flattening; and column~(5) its total
  $I$-band luminosity.  All Gaussians have the same position angle and
  the same center.}
\end{deluxetable}

%%%%%%%%%%%%%%%
% End of Document
%%%%%%%%%%%%%%%

\end{document}